\documentclass[aps,pra,reprint,groupedaddress,twocolumn,longbibliography]{revtex4-2}
\usepackage{graphicx}
\graphicspath{{figures/}}
\usepackage[dvipsnames]{xcolor}
\definecolor{DarkGreen}{rgb}{0.0, 0.5, 0.0}
\usepackage[colorlinks,citecolor=DarkGreen,linkcolor=blue,linktocpage]{hyperref}

\usepackage[T1]{fontenc}
\usepackage{textcomp}
\usepackage{bm}
\usepackage{mathtools}
\usepackage{amsmath,amssymb,amsthm}
\usepackage{comment}
\usepackage{grffile}

\usepackage{braket}        
\usepackage{booktabs}      

\newcommand{\Mprime}{M'}
\newcommand{\Ztot}{Z(l)}
\newcommand{\Dicke}[2]{\ket{D_{#1}^{#2}}}
\newcommand{\AQMBS}[2]{\ket{{AS}_{#1}(#2)}}

\newcommand{\half}{\tfrac12}
\newcommand{\Ham}{\mathcal{H}}

\newcommand{\figpanel}[2]{%
  \sbox0{#2}%
  \makebox[\columnwidth][s]{%
    \rlap{\raisebox{\dimexpr\ht0-\height\relax}[0pt][0pt]{\small #1}}%
    \hfill\usebox0\hfill}}
\newcommand{\circuitpdf}[2][0]{\raisebox{-#1\height}{#2}}

\extrafloats{100}
\newcommand{\smref}[2]{#2}

\begin{document}


\title{Unified preparation of exact and asymptotic quantum many-body scars from Dicke states}

\author{Yuki Asari}
\email{1225503@ed.tus.ac.jp}
\affiliation{Department of Physics, Tokyo University of Science, 1-3 Kagurazaka, Tokyo 162-8601, Japan}
\author{Masaya Kunimi}
\email{kunimi@rs.tus.ac.jp}
\affiliation{Department of Physics, Tokyo University of Science, 1-3 Kagurazaka, Tokyo 162-8601, Japan}
\author{Tetsuro Nikuni}
\email{nikuni@rs.tus.ac.jp}
\affiliation{Department of Physics, Tokyo University of Science, 1-3 Kagurazaka, Tokyo 162-8601, Japan}


\date{\today}

\begin{abstract}
We propose a unified quantum algorithm for preparing exact and asymptotic quantum many-body scar (QMBS) states in the spin-1 {\it XY}, domain-wall-conserving, and Onsager scar models. Model-dependent isometries map the exact scar states to Dicke states and the target asymptotic QMBS states to gapless magnon excitations of the spin-1/2 ferromagnetic Heisenberg chain. Exact QMBS states are prepared by applying the inverse isometries to Dicke states. For asymptotic QMBS states, block encoding first prepares the magnon excitations from Dicke states to a prescribed fidelity, after which the inverse isometries map the prepared states back to the physical chains. At fixed target fidelity, the block-encoding success probability has a positive lower bound independent of system size and scar index, implying that a constant number of amplitude amplification rounds suffices. Our algorithm enables systematic studies of nonergodic dynamics on programmable quantum simulators.
\end{abstract}
\maketitle

\emph{Introduction.} Thermalization of isolated quantum many-body systems is a cornerstone of quantum statistical physics. Studied by von Neumann~\cite{vonNeumann1929}, thermalization is now commonly understood in terms of the eigenstate thermalization hypothesis (ETH)~\cite{Deutsch1991,Srednicki1994,RigolDunjkoOlshanii2008,DAlessio2016}, with experimental advances across isolated quantum platforms---ultracold atoms~\cite{Kinoshita2006,Trotzky2012,Kaufman2016}, trapped ions~\cite{Clos2016}, and superconducting processors~\cite{Neill2016,Andersen2025}. Nonintegrable systems are typically expected to satisfy the ETH~\cite{BiroliKollathLauchli2010,DAlessio2016}, which thus provides a widely applicable equilibration paradigm.

Quantum many-body scar (QMBS) states, which are nonthermal eigenstates coexisting with an otherwise thermal spectrum~\cite{SerbynReview2021,MoudgalyaReview2022,ChandranReview2023}, are notable exceptions to this paradigm. Long-lived revivals observed in a Rydberg array~\cite{Bernien2017} were subsequently linked to atypical energy eigenstates of the PXP model~\cite{Turner2018a,Turner2018b,Ren2024} whose Hilbert space is constrained by the Rydberg blockade~\cite{LesanovskyKatsura2012}. The violation of the ETH by QMBS states is understood through several approaches. One approach is given by Shiraishi and Mori~\cite{ShiraishiMori2017}, which embeds nonthermal states using local projectors. Another approach is given by Mark {\it et al.}~\cite{Mark2020} and Moudgalya {\it et al.}~\cite{MoudgalyaEtaPairing2020}, which is based on a spectrum generating algebra restricted to the scarred subspace. Related work has also investigated quasi-Nambu--Goldstone modes associated with quantum many-body scar towers~\cite{Ren2024}. The QMBS states have been studied extensively, both theoretically~\cite{SchecterIadecola2019,IadecolaSchecter2020,ShibataYoshiokaKatsura2020} and experimentally~\cite{Bernien2017,Bluvstein2021,Su2023,ChenBurdick2022,ZhangDong2023,DongDesaules2023,ZhaoLoh2025}.

Gotta {\it et al.}\ proposed asymptotic QMBS (AQMBS) states~\cite{Gotta2023}, which exhibit three distinct properties: (i)~orthogonality to all exact QMBS states, (ii)~low entanglement, and (iii)~vanishing energy variance $\Delta E^{2}$ in the thermodynamic limit. According to the Mandelstam--Tamm bound~\cite{MandelstamTamm1945}, the orthogonalization time satisfies $\tau_{\mathrm{ortho}}\ge\pi\hbar/(2\Delta E)$. Together with property~(iii), the bound implies that the orthogonalization time diverges with the system size. Kunimi \emph{et al.} developed a parent-Hamiltonian approach to systematic construction of AQMBS states~\cite{Kunimi2025}. In that work, the AQMBS states of the spin-1 {\it XY} model~\cite{SchecterIadecola2019}, the domain-wall-conserving (DWC) model~\cite{IadecolaSchecter2020}, and the Onsager scar model~\cite{ShibataYoshiokaKatsura2020,Tamura2022quantum} were constructed as excitations of their parent Hamiltonians. Recently, this approach has been extended to systems with higher symmetry, such as SU($N$) Hubbard models~\cite{Hashimoto2026}. Furthermore, recent studies have linked AQMBS to diffusive Nambu--Goldstone modes associated with strong-to-weak spontaneous symmetry breaking~\cite{Kunimi2026correspondence} and to gapless excitations at quantum critical points~\cite{Logaric2026phasetransitions}. AQMBS states have also been established for broad classes of local Hamiltonians admitting the $W$ state as an exact scar~\cite{Gioia2025distinct}.

To observe nonergodic signatures in quantum simulators~\cite{CiracZoller2012,Altman2021,Daley2022,Fauseweh2024,Bloch2008,GrossBloch2017,BlattRoos2012,Monroe2021,BrowaeysLahaye2020,AspuruGuzikWalther2012,Carusotto2020,King2025}, we need to prepare the initial state appropriately. 
Two common approaches are adiabatic and digital state preparation. The adiabatic approach~\cite{Farhi2000,AlbashLidar2018} is realized by detuning sweeps in Rydberg arrays~\cite{Bernien2017,Bluvstein2021} and lattice loading and rearrangement protocols in a Bose--Hubbard simulator~\cite{Su2023,Honda2025observation}.  Digital preparation~\cite{Lloyd1996,NielsenChuang2010,Preskill2018} uses sequences of quantum gates implemented by control pulses on superconducting processors~\cite{ChenBurdick2022,ZhangDong2023,DongDesaules2023} and Rydberg arrays~\cite{ZhaoLoh2025}. Because the target AQMBS states correspond to gapless excitations of their parent Hamiltonians~\cite{Kunimi2025}, direct adiabatic preparation can be challenging, whereas the digital approach can prepare any quantum state in principle~\cite{NielsenChuang2010}. More recently, analog-digital hybrid quantum simulation---in which state preparation is digital and time evolution is analog---has been realized on neutral-atom arrays~\cite{Geim2026, Timme2026} and on a superconducting processor~\cite{Andersen2025}. Thus, one approach to observing these AQMBS signatures is to prepare the state digitally and evolve it using either digital circuits or analog dynamics.

Existing approaches include digital preparation of exact QMBS states~\cite{Gustafson2023,Buhrman2024, Guo2026}, variational preparation using shallow circuits~\cite{Cenedese2025shallow,Carolan2026variational}, and preparation of a single-magnon AQMBS state on a trapped-ion processor in logarithmic circuit depth~\cite{Logaric2026}.

In this Letter, we present a unified preparation approach based on the fact that AQMBS states in several models correspond to magnon excited states of the spin-1/2 ferromagnetic Heisenberg model, which appears as the parent Hamiltonian used in the AQMBS construction~\cite{Kunimi2025}. Our approach is based on quantum circuits and uses block encoding~\cite{LowChuang2017,LowChuang2019,Gilyen2019} to implement a nonunitary operator that transforms the Dicke state, which is a ground state of the spin-1/2 ferromagnetic Heisenberg model, into the magnon excited state. The model-dependent mapping of our approach consists of operations that (i) increase the number of qubits and (ii) rearrange the excitations into configurations allowed by the model up to a configuration-dependent sign. We treat the spin-1 {\it XY} model in the main text and the DWC and Onsager scar models in the Supplemental Material (SM)~\footnote{See Supplemental Material at [URL], which includes Refs.~\cite{Hoeffding1963,Williams1991,Barenco1995,KhattarGidney2025,Fowler2012,Nielsen2002,Bluvstein2024,Evered2026,QuantinuumH1,IBMHeron,Yauk2024,BianchiDona2019,Page1993}, for the choice of $\tau$ and its guarantees, the block-encoding circuit, the domain-wall-conserving and Onsager constructions, the noise-robustness analysis, and the entanglement of the prepared state.}.

{\it Preparation of magnon excited states.} We now present the details of our approach for preparing the magnon excited states. First, we discuss the first-order restricted spectrum generating algebra (RSGA-1)~\cite{MoudgalyaEtaPairing2020,Mark2020,Batista2009} that underlies the AQMBS states we consider. Let us consider a nonintegrable Hamiltonian $\Ham$, a normalized state $\ket{S_0}$, and a ladder operator $Q^\dagger$ satisfying
\begin{subequations}\label{eq:rsga}
\begin{align}
  &\Ham\ket{S_0}=E_0\ket{S_0},\\
  &[\Ham, Q^\dagger]\ket{S_0}=\mathcal{E} Q^\dagger\ket{S_0},
    \qquad Q^\dagger\ket{S_0}\neq0,\\
  &\bigl[[\Ham, Q^\dagger], Q^\dagger\bigr]=0,\\
  & Q^\dagger=\sum_{j=1}^{M} q^{\dagger}_j,\label{eq:rsga_local}
\end{align}
\end{subequations}
where $E_0$ and $\mathcal{E}$ are real numbers, $M$ is the total number of lattice sites, and $ q^{\dagger}_j$ are the local terms of the non-Hermitian ladder operator. From these conditions, we can show that the following relations hold:
\begin{align}
\Ham\ket{S_n}=(E_0+n\mathcal{E})\ket{S_n},\quad \ket{S_n}\propto(Q^{\dagger})^n\ket{S_0},\label{eq:eigenequations_for_scar_states}
\end{align}
where $\ket{S_n}$ is a QMBS state and we refer to $n$ as the scar index.

For the spin-1 {\it XY} model introduced below, the corresponding AQMBS state $\AQMBS{n}{l}$ is obtained by replacing one of the operators as follows~\cite{Gotta2023,Moudgalya2024symmetries,Kunimi2025}:
\begin{align}
  \AQMBS{n}{l}&\propto Q^{\dagger}(l)( Q^{\dagger})^{n-1}\ket{S_0},\label{eq:aqmbs_tower}\\
   Q^\dagger(l)&\equiv\sum_{j=1}^{M}f_j(l) q^{\dagger}_j,\label{eq:definition_of_Q_dagger_of_l}\\
  f_j(l)&\equiv\cos\!\left[\frac{\pi l(2j-1)}{2M}\right],\;l=1,2,\ldots,M-1.\label{eq:definition_of_f_j_l}
\end{align}
Here, we assume open boundary conditions. The energy variance $\Delta E^{2}\equiv\braket{\Ham^{2}}-\braket{\Ham}^{2}$ obtained from $\AQMBS{n}{l}$ vanishes in the thermodynamic limit when $l$ is $O(1)$~\cite{Gotta2023,Kunimi2025}. As shown in Ref.~\cite{Kunimi2024}, AQMBS states are not unique. 
Alternative modulation coefficients can also be used, provided the resulting states remain orthogonal to the exact scar tower and have vanishing energy variance in the thermodynamic limit. For some models discussed below, the ideal AQMBS states are constructed on the effective chain and mapped back to the physical chain, with the modulation coefficients evaluated in the effective coordinates.

\begin{figure}[tb]
\centering
\figpanel{(a)}{\circuitpdf[0.45758]{\includegraphics[hiresbb,width=0.74\columnwidth]{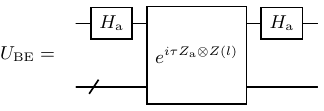}}}\\[4pt]
\figpanel{(b)}{\circuitpdf[0.58763]{\includegraphics[hiresbb,width=\columnwidth]{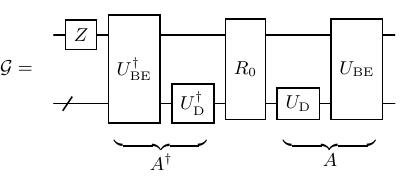}}}\\[4pt]
\figpanel{(c)}{\circuitpdf[0.45486]{\includegraphics[hiresbb,width=\columnwidth]{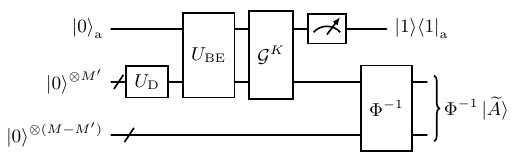}}}
\caption{Components of the preparation circuit. (a)~The block-encoding unitary $U_{\mathrm{BE}}$. (b)~The Grover iterate $\mathcal{G}=AR_0A^{\dagger}Z_{\mathrm{a}}$ with $A=U_{\mathrm{BE}}( I_{\mathrm{a}}\otimes U_{\mathrm{D}})$ and $R_0=2\ket{0\cdots0}\!\bra{0\cdots0}-I$, the reflection about the all-zero state of the ancilla and the data qubits. (c)~Full pipeline, where $U_{\mathrm{D}}$ is the Dicke preparation circuit~\cite{BartschiEidenbenz2019} and $\Phi^{-1}=\mathcal D\circ\mathcal S\circ\mathcal R$ is the isometry that maps the magnon excited states to the corresponding AQMBS states. The circuit prepares $\Phi^{-1}|\tilde{A}\rangle$, where $|\widetilde{A}\rangle$ is defined in Eq.~(S1). This output approximates the target AQMBS state $|AS_n(l)\rangle$. }
\label{fig:circuit}
\end{figure}

As an example of a system possessing QMBS and AQMBS states, we consider the spin-1 {\it XY} model~\cite{SchecterIadecola2019,Gotta2023,Kunimi2025},
\begin{align}\label{eq:spin1xy}
  \Ham&=J\sum_{j=1}^{M-1}( \tau^x_j \tau^x_{j+1}+ \tau^y_j \tau^y_{j+1})
   +h\sum_{j=1}^M \tau^z_j+D\sum_{j=1}^M( \tau^z_j)^2\nonumber\\
   &\quad+J_3\sum_{j=1}^{M-3}( \tau^x_j \tau^x_{j+3}+ \tau^y_j \tau^y_{j+3}),
\end{align}
where $ \tau^{x,y,z}_j$ are spin-1 operators at site $j$ and $J$, $h$, $D$, and $J_3$ are coupling constants. The ladder operator and the state $\ket{S_0}$ are defined as~\cite{SchecterIadecola2019}
\begin{align}
   Q^\dagger&\equiv\frac{1}{2}\sum_{j=1}^{M}(-1)^{j}( \tau^+_j)^{2},\\
  \ket{S_0}&\equiv\ket{-_1,-_2,\ldots,-_M},
\end{align}
where $\tau^\pm_j\equiv \tau^x_j\pm i \tau^y_j$ are spin-1 ladder operators. Here, $\ket{+_j}$, $\ket{0_j}$, and $\ket{-_j}$ are the eigenstates of $ \tau^z_j$ with eigenvalues $+1$, $0$, and $-1$, respectively. We can show that $\Ham$, $ Q^\dagger$, and $\ket{S_0}$ satisfy RSGA-1~\cite{MoudgalyaEtaPairing2020}. The local terms of $ Q^\dagger$ are $ q^{\dagger}_j=\frac{1}{2}(-1)^{j}( \tau^+_j)^{2}$.

\begin{figure*}[!t]
\centering
\includegraphics[width=\textwidth]{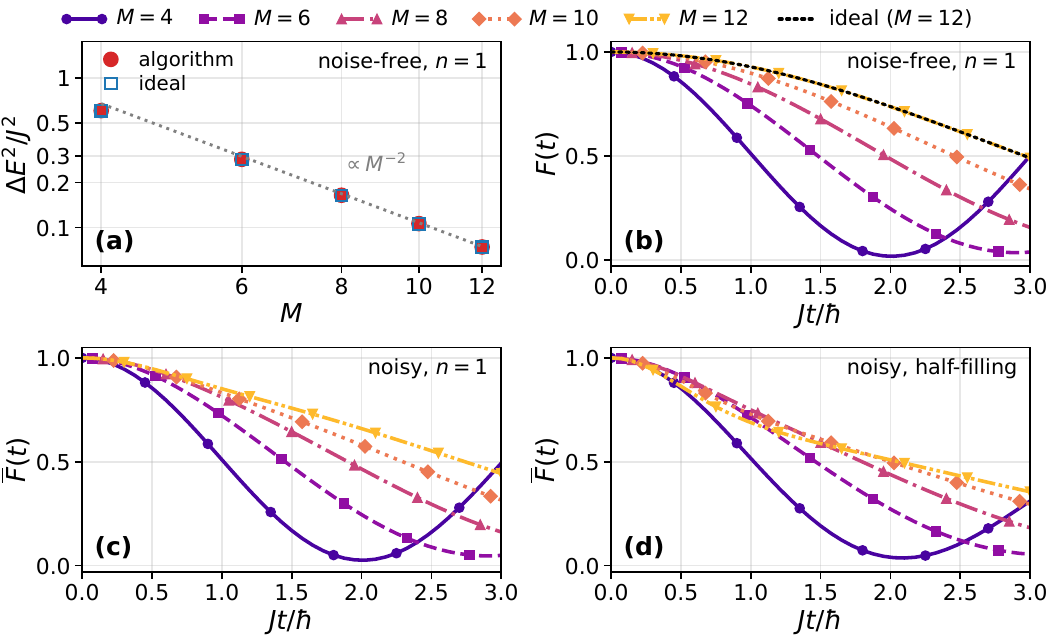}
\caption{Algorithm-prepared AQMBS state of the spin-1 {\it XY} model~\eqref{eq:spin1xy} for $h/J=0$, $D/J=0.1$, and $J_3/J=0.1$. (a) The red circles represent the energy variance $\Delta E^2/J^2$ calculated from the algorithm-prepared state and the blue squares represent the energy variance calculated from Eq.~(\ref{eq:aqmbs_tower}) for $n=1$ and $l=1$. The gray dotted line represents the analytical results for large $M$ [$\Delta E^2\simeq (\pi^2l^2/M^2)(J^2+9J_3^2)$]. (b) Time evolution of the fidelity $F(t)=|\braket{\psi(0)|\psi(t)}|^{2}$ for $M=4, 6, 8, 10$ and $12$. The initial state is an algorithm-prepared AQMBS state for $n=1$ and $l=1$ without noise. The black dashed line shows the fidelity for the ideal AQMBS state in Eq.~(\ref{eq:aqmbs_tower}) at $M=12$. (c) The noise-averaged fidelity. Noise is included in the state preparation stage. The error rates of one- and two-qubit gates are $10^{-4}$ and $1.5\times 10^{-3}$, respectively. We do not include errors in the time evolution. (d) The time evolution of the noise-averaged fidelity for $n=M/2$.
}
\label{fig:relax}
\end{figure*}

Direct calculations show that the parent Hamiltonian of the QMBS state $\ket{S_n}$ becomes the spin-1/2 ferromagnetic Heisenberg model after an appropriate unitary transformation~\cite{Kunimi2025}. In this correspondence, we assign the labels $\ket{+}=\ket{1}$ and $\ket{-}=\ket{0}$ to an effective two-level system~\footnote{With this identification, the spin-1/2 operators can be written as $ S^z=(1/2)\bigl(\ket{1}\!\bra{1}-\ket{0}\!\bra{0}\bigr)=-(1/2)Z$ and $ S^+\equiv\ket{1}\!\bra{0}$.}. The QMBS state of the spin-1 {\it XY} model can be expressed as the Dicke state
\begin{equation}\label{eq:dicke}
  \Dicke{n}{M}=\binom{M}{n}^{-1/2}\sum_{|T|=n}\ket{T},
\end{equation}
where $\ket{T}$ denotes a computational basis state with $n$ excitations, so that $\Dicke{n}{M}$ is the equal superposition of all $\binom{M}{n}$ such states. For $J_3=0$, the profile in Eq.~(\ref{eq:definition_of_f_j_l}) is an exact magnon eigenmode of the open nearest-neighbor ferromagnetic Heisenberg chain~\cite{Kunimi2025}.
This state is obtained by acting on the Dicke state with the following nonunitary operator~\cite{Kunimi2026correspondence}:
\begin{equation}\label{eq:filter}
    \Ztot\equiv\sum_{j=1}^{M}f_j(l) Z_j,
\end{equation}
where $Z_j$ is the Pauli-$Z$ operator at site $j$. We use the same profile for $J_3\neq0$ in Fig.~\ref{fig:relax}, where the energy variance of the ideal target state still scales as $M^{-2}$ at fixed $l$.

From the above results, our strategy for preparing AQMBS states in quantum circuits is as follows: We prepare the Dicke state and then apply the nonunitary operator $\Ztot$ by block encoding~\cite{LowChuang2017,LowChuang2019,Gilyen2019}. We initialize $M$ qubits and one ancilla in $\ket{0}^{\otimes (M+1)}$, and then convert the $M$ qubits into the Dicke state $\Dicke{n}{M}$. Dicke states can be prepared using various approaches~\cite{BartschiEidenbenz2019,BartschiEidenbenz2022,Aktar2022,JoshiVasconcelos2026,LiuChildsGottesman2024,VanDyke2021,Piroli2024,Buhrman2024,Yu2026,Nagib2025,NepomechieRaveh2024,RavehNepomechie2024,NepomechieRavaniniRaveh2024,Chen2026,Mukherjee2020,Farrell2025,PannierGunther2026,Haffner2005,Hume2009}. Among them, the B{\"a}rtschi--Eidenbenz protocol prepares the Dicke state with an ancilla-free circuit of depth $O(M)$ requiring only linear-nearest-neighbor connectivity~\cite{BartschiEidenbenz2019}. The depth can be reduced to $O(n\log(M/n))$ for all-to-all and $O(\sqrt{Mn})$ for grid connectivity~\cite{BartschiEidenbenz2022}. The circuit diagram for the block encoding is shown in Fig.~\ref{fig:circuit}. Running the circuit of Fig.~\ref{fig:circuit}(a) and postselecting the ancilla on $\ket{1}$ then yields~\cite{Roggero2019dynamic,Roggero2020preparation}
\begin{equation}\label{eq:sin_dicke}
  \bra{1}_{\mathrm{a}} H_{\mathrm{a}} e^{i\tau Z_{\mathrm{a}}\otimes\Ztot} H_{\mathrm{a}} \ket{0}_{\mathrm{a}}\Dicke{n}{M}
  = i\sin[\tau\Ztot]\Dicke{n}{M} ,
\end{equation}
where $H_{\mathrm{a}}$ and $Z_{\mathrm{a}}$ are the Hadamard gate and the Pauli-$Z$ operator on the ancilla, and $\tau$ is a tunable angle. The expectation value of $\Ztot$ with respect to the Dicke state vanishes because $\sum_{j=1}^Mf_j(l)=0$ for $l=1,2,\ldots,M-1$. From this relation, we obtain
$\braket{\Ztot^{2}}=2n(M-n)/(M-1)$ [see details in SM A.1]. For the numerical simulations, we choose
\begin{equation}\label{eq:tau_choice}
  \tau=\frac{s}{\sqrt{\braket{\Ztot^{2}}}},\qquad s\equiv 0.495.
\end{equation}
This choice is motivated by the Gaussian-limit expressions in SM A.1. Numerical evaluations give an infidelity $1-F\leq0.01$ between the normalized states proportional to $\sin[\tau Z(l)]|D_n^M\rangle$ and $Z(l)|D_n^M\rangle$ for the parameter sets examined. In the large-$M$ limit at fixed filling, the success probability approaches
\begin{align}
p_{\mathrm{succ}}\simeq\frac{1}{2}\left(1-e^{-2s^2}\right)\approx0.19.\label{eq:p_success}
\end{align}
Finite-size evaluations give $0.19\leq p_{\mathrm{succ}}\leq0.23$ over the parameter sets examined (see SM A.1).

To increase the success probability to near unity, we apply amplitude amplification to the branch with the ancilla in $\ket{1}_{\mathrm{a}}$~\cite{Grover1996,Brassard2002}.
Each round applies the Grover iteration $\mathcal{G}$ shown in Fig.~\ref{fig:circuit}(b), and the number of rounds required is
\begin{equation}\label{eq:K_period}
  K=\left\lfloor\frac{\pi}{4\arcsin(\sqrt{p_{\mathrm{succ}}})}\right\rfloor,
\end{equation}
the standard choice of Ref.~\cite{Brassard2002}. The success probability admits a positive lower bound independent of $n$ and $M$, ensuring $K=O(1)$ (see SM A.1). For $0.19\leq p_{\mathrm{succ}}\leq0.23$, a single amplification round raises the success probability above 0.95.

\emph{Model-dependent mapping.} The remaining two models are the DWC model~\cite{IadecolaSchecter2020} and the Onsager scar model~\cite{ShibataYoshiokaKatsura2020}, whose Hamiltonians are defined as
\begin{align}
  \Ham_{\mathrm{DWC}}={}&J\sum_{j=1}^{M} S^x_j\bigl(1-4 S^z_{j-1} S^z_{j+1}\bigr)\nonumber\\
  &-h\sum_{j=1}^{M} S^z_j+J_z\sum_{j=0}^{M} S^z_j S^z_{j+1},\label{eq:hdwc}\\
  \Ham_{\mathrm{Ons}}={}&-2J\sum_{j=1}^{M-1}\bigl( S^x_j S^x_{j+1}+ S^y_j S^y_{j+1}\bigr)\nonumber\\
  &+J_{\mathrm{c}}\sum_{j=1}^{M} h_{{\mathrm{c}},j}-h\sum_{j=1}^{M} S^z_j,\label{eq:hons}\\
  h_{{\mathrm{c}},j}={}&\frac12\bigl( P_{j-1} P'_j+ P'_j P_{j+1}\nonumber\\
  &+ S^-_{j-1} P'_j S^+_{j+1}+ S^+_{j-1} P'_j S^-_{j+1}\bigr),\label{eq:hc}
\end{align}
where $ S^{x,y,z}_j$ are spin-$1/2$ operators at site $j$, $ S^\pm_j\equiv S^x_j\pm i S^y_j$, $ P_j\equiv\half- S^z_j$ and $ P'_j\equiv\half+ S^z_j$ project onto $\ket{0}_j$ and $\ket{1}_j$, respectively, and $J$, $h$, $J_z$, and $J_{\mathrm{c}}$ are coupling constants. The boundary conventions are $ S^z_0= S^z_{M+1}\equiv-\half$ for the DWC model, and $ P_0= P_{M+1}= S^\pm_0= S^\pm_{M+1}=0$ for the Onsager scar model. The ladder operators for generating QMBS states are defined as
\begin{align}
   Q^\dagger_{\mathrm{DWC}}&\equiv\sum_{j=1}^{M}(-1)^{j} P_{j-1} S^+_j P_{j+1},\label{eq:qdwc}\\
   Q^\dagger_{\mathrm{Ons}}&\equiv\sum_{j=1}^{M-1}(-1)^{j-1} S^+_j S^+_{j+1} .\label{eq:qons}
\end{align}
The QMBS states for both models are built on the fully polarized state $\ket{S_0}=\ket{0}^{\otimes M}$. 
The excitation patterns for the QMBS states are restricted due to the functional form of the ladder operators: the projectors of $Q^\dagger_{\mathrm{DWC}}$ forbid adjacent excitations, whereas $Q^\dagger_{\mathrm{Ons}}$ creates excitations only as adjacent pairs.

As shown in Ref.~\cite{Kunimi2025}, the parent Hamiltonians for the above models are different. However, these parent Hamiltonians can be mapped to the spin-1/2 ferromagnetic Heisenberg model with $M'\le M$ lattice sites  ~\cite{GomezSantos1993,CheongHenley2009,Kunimi2025}.

Crucially, the inverse mapping is implementable as an isometry. The circuit $\Phi^{-1}$~\footnote{
For fixed $n$, $\Phi$ is an isometry from the constrained sector onto the weight-$n$ subspace of the effective chain; its inverse is well defined on this subspace. The circuit extends $\Phi^{-1}$ to the whole effective-chain Hilbert space. With $\Phi=(\Phi^{-1})^\dagger$ for this extension, $\Phi\Phi^{-1}=I$ on the effective chain, whereas $\Phi^{-1}\Phi$ projects onto the image of the extended circuit.} relabels and locally dresses basis configurations while increasing the number of qubits to $M$. It maps each basis configuration to a single signed basis configuration rather than to a superposition. The same preparation circuit thus serves all three models. The model enters only through $\Phi^{-1}$ and $\Mprime$, which is $M-n+1$ for the DWC model and $M-n$ for the Onsager scar model (see details in  SM~B)~\cite{Bariev1991,Chhajlany2016,PozsgayGomborHutsalyuk2021,BorsiPristyakPozsgay2023,BorsiPozsgay2025}. The preparation algorithm then runs on the effective chain with $M$ replaced by $\Mprime$.

\begin{table}[tb]
\caption{Resource costs of the protocol for $n=O(1)$ and $n=\Mprime/2$. Here, we adopt the B{\"a}rtschi--Eidenbenz Dicke circuit for the Dicke state preparation, which has circuit depth $\Theta(\Mprime)$ and uses $\Theta(n\Mprime)$ gates~~\cite{BartschiEidenbenz2019}. The infidelity bound for the choice (\ref{eq:tau_choice}) is numerically verified over the parameter range specified in SM A.1; rigorous uniform guarantees for a more conservative choice of $\tau$ are given there. 
We note that the cost of $\Phi^{-1}$ is not included in the list.
}

\label{tab:cost}
\begin{ruledtabular}
\begin{tabular}{lcc}
 & $n=O(1)$ & $n=\Mprime/2$ \\
\colrule
Grover rounds $K$ & $O(1)$ & $O(1)$ \\
Success probability $p_{\mathrm{succ}}$ & $\Theta(1)$ & $\Theta(1)$ \\
Effective-chain depth & $\Theta(\Mprime)$ & $\Theta(\Mprime)$ \\
Effective-chain gates & $\Theta(\Mprime)$ & $\Theta(\Mprime^{2})$ \\
Qubits $M$ (Spin-1 {\it XY}) & $\Mprime$ & $\Mprime$ \\
Qubits $M$ (DWC and Onsager)& $\Mprime+O(1)$ & $\tfrac{3}{2}\Mprime+O(1)$ \\
Ancillas (flag $+$ trees) & $1+\Theta(\Mprime)$ & $1+\Theta(\Mprime)$ \\
Infidelity $1-F$ & $\le 0.01$ & $\le 0.01$ \\
\end{tabular}
\end{ruledtabular}
\end{table}

The isometry $\Phi^{-1}$ is factorized into three primitives,
\begin{equation}\label{eq:rsd}
  \Phi^{-1}=\mathcal{D}\circ\mathcal{S}\circ\mathcal{R}.
\end{equation}
We number the $n$ excitations of an effective basis configuration from the left, starting at $0$, and write $q_k$ for the site of the $k$-th excitation, so that $q_0<\dots<q_{n-1}$. (i) $\mathcal{R}$ moves each excitation to the right by its own index, $\ket{1}_{q_k}\to\ket{1}_{q_k+k}$; the leftmost excitation stays put, and neighboring excitations end up at least two sites apart. During this process, the $M-M'$ qubits in $\ket{0}$ are appended. This operation can be implemented by using Fredkin gates. (ii) $\mathcal{S}$ adds the staggered phase $(-1)^{\sum_k(q_k+k)}$ inherited from the ladder operators. (iii) $\mathcal{D}$ generates the pair excitation $\ket{1}\to\ket{11}$. This can be implemented by CNOT gates. Operations $\mathcal{R}$ and $\mathcal{S}$ are necessary for the DWC and Onsager scar models, and $\mathcal{D}$ for the Onsager scar model (see details in SM~B). With the block-encoding stage omitted, the same circuit prepares the exact scar states $\ket{S_n}$.

\emph{Results of numerical simulations: spin-1 {\it XY} model.} We numerically simulate the  protocol on the spin-1 {\it XY} model~\eqref{eq:spin1xy}, which hosts the scar tower generated by $Q^{\dagger}$ on the fully polarized state $\ket{S_0}$. For simplicity, we consider the case of even $M$. In this case, the isometry $\Phi^{-1}$ reduces to $\prod_{j=1}^{M/2}Z_{2j-1}$, which means that no additional qubits are required. Each spin-1 site becomes one site of the effective spin-1/2 chain, spanned by $\ket{+}_j$ and $\ket{-}_j$. The algorithm comprises Dicke state preparation, the block encoding of $\Ztot$, and amplitude amplification on $M$ qubits. With the choice~\eqref{eq:tau_choice}, the success probability is $p_{\mathrm{succ}}\simeq 0.22$ at $n=1$ and $p_{\mathrm{succ}}\simeq 0.20$ at half filling ($n=M/2$) over the sizes studied (see details in SM~A.1). Thus, neither case requires more than one amplification round.

For $n=1$ and $l=1$, Fig.~\ref{fig:relax}(a) compares the energy variances of the algorithm-prepared state and the ideal state in Eq.~(\ref{eq:aqmbs_tower}) with the large-$M$ expression $\Delta E^2\simeq(\pi^2l^2/M^2)(J^2+9J_3^2)$~\cite{Gotta2023}.
Figure~\ref{fig:relax}(b) shows the time evolution of the fidelity $F(t)\equiv |\langle\psi(0)|\psi(t)\rangle|^2$ under the Hamiltonian (\ref{eq:spin1xy}), where the initial state is the AQMBS state constructed by our algorithm. As expected, the short-time behavior of the fidelity is consistent with the $M^{-2}$ scaling of the energy variance.

We also perform numerical simulations including gate errors to clarify the noise robustness. Figures~\ref{fig:relax}(c) and \ref{fig:relax}(d) show the time evolution of the noise-averaged fidelity $\overline{F}(t)\equiv \sum_j p_{\mathrm{succ}, j} |\langle\psi_j(0)|e^{-i\mathcal{H}t/\hbar}|\psi_j(0)\rangle|^2/ \sum_{j} p_{\mathrm{succ},j }$, where $|\psi_j(0)\rangle$ is the postselected initial state of the $j$th trajectory and $p_{\mathrm{succ}, j}$ is its success probability. This quantity is the trajectory-averaged survival probability for the specified Pauli-error unraveling and is used as a diagnostic of the simulated trajectories. We note that $\overline{F}(t)$ is weighted by $p_{\mathrm{succ},j}$ rather than averaged uniformly over trajectories, since the $j$th trajectory passes postselection with probability $p_{\mathrm{succ},j}$. This weighting thus reduces statistical fluctuations arising from stochastic postselection. 
Gate errors are included only during state preparation.
For these simulations, we prepare the initial states without amplification ($K=0$) and compute 2400 quantum trajectories for each data point. See details in SM~C. For $n=1$, the initial decay of $\overline{F}(t)$ becomes slower with increasing $M$, even with gate errors during state preparation [Fig.~\ref{fig:relax}(c)]. At half filling, the $M=12$ curve lies below those for $M=6$--$10$ at short times [Fig.~\ref{fig:relax}(d)]. For $Jt/\hbar\gtrsim2$, however, $\overline{F}(t)$ increases with $M$ for $M=6$--$12$. This initial reversal is consistent with greater error accumulation in the $\Theta(M^2)$-gate Dicke preparation circuit $U_{\mathrm{D}}$ (Table~\ref{tab:cost}). Supplemental Material B gives the analogous results for the DWC and Onsager scar models, and SM~\textcolor{blue}{C} the systematic dependence on the one- and two-qubit error rates. 

Finally, we discuss the half-chain entanglement entropy (EE) of the AQMBS state. 
The AQMBS states considered here are expected to exhibit subvolume-law entanglement.
At half filling, we verify that the algorithmically prepared AQMBS states exhibit logarithmic scaling, $S\simeq\frac{1}{2}\ln M$, in contrast to the volume-law scaling of the Page value for Haar-random states. The results are shown in SM D.

\emph{Discussion and outlook.} The algorithm proposed in this Letter is not restricted to the three models treated here. 
The same pipeline applies whenever a scar tower admits an efficiently implementable mapping to the Dicke tower with the required operator correspondence.
Examples include towers generated by on-site ladder operators, such as those in the spin-1 {\it XY} and $\eta$-pairing models~\cite{MoudgalyaEtaPairing2020}. For these towers, $\mathcal{R}$ and $\mathcal{D}$ are trivial, and $\Phi^{-1}$ reduces to on-site relabeling and the staggered phase. The DWC and Onsager scar models have towers with constrained excitation patterns, for which $\Phi^{-1}$ appends qubits.
Our approach may extend to SU$(N)$ Hubbard models~\cite{Hashimoto2026} using qudit Dicke states~\cite{NepomechieRaveh2024} and flavor-selective diagonal operators, although implementing the required qudit gates remains nontrivial.

We have not established the required efficiently implementable mapping to Dicke states for the QMBS states of the AKLT and PXP models~\cite{Mark2020,Turner2018a}.

The choice of $\tau$ in Eq.~\eqref{eq:tau_choice} ensures that $K=O(1)$ 
for all nontrivial scar indices $1\le n\le M'-1$. Thus, amplitude amplification introduces only a constant-factor overhead. The asymptotic cost of the effective-chain stage is therefore determined by Dicke state preparation $U_{\rm D}$.
The block encoding and the reflection $R_0$ [Fig.~\ref{fig:circuit}(b)] have depth $O(\log\Mprime)$, and the routing $\Phi^{-1}$ is applied once outside the loop, whereas $U_{\mathrm{D}}$ enters each Grover iterate twice at depth $\Theta(\Mprime)$ with the B{\"a}rtschi--Eidenbenz protocol~\cite{BartschiEidenbenz2019}. 
Faster Dicke-state preparation can reduce the depth of the effective-chain stage: the symmetrization-based protocol of Ref.~\cite{LiuChildsGottesman2024} reaches polylogarithmic depth, and the protocol of Ref.~\cite{JoshiVasconcelos2026} reaches constant depth with global gates and more ancillas. Therefore, the whole effective-chain pipeline would run in polylogarithmic depth throughout this range. For the DWC and Onsager scar models, the end-to-end depth may nevertheless be set by the routing circuit $\Phi^{-1}$, which lies outside this stage (see details in SM~B.2).

A near-term demonstration of the relaxation-time divergence at $n=O(1)$, beyond the single-magnon states of Ref.~\cite{Logaric2026}, appears feasible on existing platforms. The block encoding succeeds with $\Theta(1)$ probability for all nontrivial scar indices $1\le n\le M'-1$ [Eq.~(\smref{eq:inequality_pcuss}{S15}) in SM A.1]. The prepared state has a fixed excitation number on the effective chain, which $\Phi^{-1}$ maps to a model-dependent conserved charge in the physical model. Errors that change this charge can be detected by computational-basis measurements, allowing the corresponding shots to be discarded~\cite{Bonet-Monroig2018}. 
Gate errors accumulated throughout state preparation, including routing, limit near-term implementations to small $n$ (see SM C).
The accessible system sizes and scar indices are therefore limited primarily by noise accumulated in the $\Theta(nM')$ gates of $U_{\mathrm D}$ (see Table~\ref{tab:cost}) and, for the DWC and Onsager scar models, in $\Phi^{-1}$.
For extensive $n$, these noise constraints still favor error-corrected hardware.

Our protocol provides a unified approach to preparing nonthermal initial states, enabling systematic investigations of nonergodic dynamics on programmable quantum simulators.

{\it Acknowledgments.}
Y.A. used AI-based tools (Claude Sonnet 5.5, Opus 5.5, and Fable 5; Anthropic) as aids for suggesting proof ideas, obtaining technical inequalities, constructing quantum circuits, developing numerical simulation codes, and improving the presentation.
All proofs, calculations, simulations, and conclusions were independently verified by all authors.
Y.A. is particularly grateful to Akira Tanji for improving the preparation algorithm.
Y.A. also thanks Simon J. Evered and Takafumi Tomita for helpful discussions. 
This work was supported by JSPS KAKENHI Grant No. JP25K00215 (M.K.) and JST ASPIRE No. JPMJAP24C2 (M.K.).

{\it Data availability.}
The supporting data for this Letter are openly available on Zenodo~\cite{Zenodo}.


\bibliography{references_aqmbs}

\end{document}


\onecolumngrid
\setcounter{equation}{0}
\renewcommand{\theequation}{S\arabic{equation}}
\renewcommand{\theHequation}{S\arabic{equation}}
\setcounter{table}{0}
\renewcommand{\thetable}{S\arabic{table}}
\renewcommand{\theHtable}{S\arabic{table}}
\setcounter{figure}{0}
\renewcommand{\thefigure}{S\arabic{figure}}
\renewcommand{\theHfigure}{S\arabic{figure}}
\renewcommand{\citenumfont}[1]{S#1}   
\renewcommand{\bibnumfmt}[1]{[S#1]}   

\begin{center}
  {\large\textbf{Supplemental Material for\\[3pt]
  ``Unified preparation of exact and asymptotic quantum many-body scars \\ from Dicke states''}}\\[1.1em]
  Yuki Asari, Masaya Kunimi, and Tetsuro Nikuni\\[0.4em]
  \textit{Department of Physics, Tokyo University of Science, 1-3 Kagurazaka, Tokyo 162-8601, Japan}
\end{center}
\medskip

\noindent \textbf{Sec.~A}: Algorithm on the effective chain---the choice of $\tau$ and the resulting infidelity and success probability. {\bf Sec.~B}: Details of the construction of quantum circuits for the domain-wall-conserving model and the Onsager scar model. \textbf{Sec.~C}: Noise robustness---current gate error rates, and the success and fidelity landscape versus gate error rate and amplification depth. \textbf{Sec.~D}: Entanglement properties of the prepared AQMBS states compared with the Page value.

\medskip
\noindent\emph{Notation.} The Pauli $Z$ operator at site $j$ is defined by $Z_j\equiv\ket{0_j}\bra{0_j}-\ket{1_j}\bra{1_j}$. The function $f_j(l)\equiv\cos[\pi l(2j-1)/(2\Mprime)]$ $(j=1,\ldots,M', l=1,\dots,\Mprime-1)$ obeys the zero-sum relation $\sum_jf_j(l)=0$. Weight-$n$ computational basis states are labeled by their excitation sets $T$, $|T|=n$. A global bit flip maps $n$ to $\Mprime-n$ and $\Ztot$ to $-\Ztot$. It therefore suffices to consider $1\leq n\leq M'/2$.

\section*{A.\quad Algorithm details}

\subsection*{A.1\quad Choice of $\tau$: infidelity and success probability}

We show how to prepare the target state $\ket{A}\propto\Ztot\Dicke{n}{\Mprime}$ with infidelity at most $\varepsilon$ and a success probability bounded below independently of $n$ and $\Mprime$. We define the normalized target state $\ket{A}$, the normalized postselected output $\ket{\tilde{A}}$, and their fidelity $F$ as follows:
\begin{equation}\label{eq:ASdef}
  \ket{A}\coloneqq\frac{\Ztot\Dicke{n}{\Mprime}}{\|\Ztot\Dicke{n}{\Mprime}\|},\qquad
  \ket{\tilde A}\coloneqq\frac{\sin[\tau\Ztot]\Dicke{n}{\Mprime}}{\|\sin[\tau\Ztot]\Dicke{n}{\Mprime}\|},\qquad
  F\coloneqq\bigl|\braket{\tilde A|A}\bigr|^{2}.
\end{equation}
The success probability is defined by $p_{\mathrm{succ}}=\|\sin[\tau\Ztot]\Dicke{n}{\Mprime}\|^{2}$. The requirement on $\tau$ is then
\begin{equation}\label{eq:target}
  1-F\le\varepsilon ,
\end{equation}
with $p_{\mathrm{succ}}$ bounded below uniformly in $n$. Since $\sin[\tau Z(l)]\simeq\tau Z(l)$ as $\tau\to0$, decreasing $\tau$ improves the fidelity but reduces the success probability $p_{\mathrm{succ}}$. The relevant scale for this trade-off is the root-mean-square value of $Z(l)$ in the Dicke state, rather than its operator norm. We will discuss this below.

The moments of $Z(l)$ in the Dicke state can be expressed as sums over computational-basis configurations. By the zero-sum relation, $\Ztot\ket T=\bigl(\sum_{j\notin T}f_j(l)-\sum_{j\in T}f_j(l)\bigr)\ket T=-2\sum_{j\in T}f_j(l)\ket T$, and $\Dicke{n}{\Mprime}$ is the equal superposition of the $\binom{\Mprime}{n}$ states $\ket T$, so that
\begin{equation}\label{eq:config}
  \braket{\Ztot^{k}}=\frac{(-2)^{k}}{\binom{\Mprime}{n}}\sum_{|T|=n}\Bigl(\sum_{j\in T}f_j(l)\Bigr)^{k}.
\end{equation}

The moment bounds require the power sums $\sum_{j=1}^{M'}f_j(l)^{k}$ for $k\le6$. We obtain them from the cosine sums $S_k(l)\equiv\sum_{j=1}^{\Mprime}\cos[k\pi l(2j-1)/(2\Mprime)]$: by the geometric series, $S_k(l)=(-1)^{q}\Mprime$ when $kl/(2\Mprime)$ is an integer $q$, and $S_k(l)=0$ otherwise. Hence $\sum_{j=1}^{M'}f_j(l)=S_1=0$, $\sum_{j=1}^{M'}f_j(l)^{2}=\Mprime/2+S_2/2=\Mprime/2$, and
\begin{equation}\label{eq:powersums}
\begin{aligned}
  \sum_{j=1}^{M'}f_j(l)^{3}&=\frac{3S_1+S_3}{4}=\begin{cases}-\Mprime/4,&3l=2\Mprime,\\0,&\text{otherwise},\end{cases}\\
  \sum_{j=1}^{M'}f_j(l)^{4}&=\frac{3\Mprime+4S_2+S_4}{8}=\begin{cases}\Mprime/4,&4l=2\Mprime,\\3\Mprime/8,&\text{otherwise},\end{cases}\\
  \sum_{j=1}^{M'}f_j(l)^{6}&=\frac{10\Mprime+15S_2+6S_4+S_6}{32}=\begin{cases}\Mprime/8,&4l=2\Mprime,\\9\Mprime/32,&6l=2\Mprime,\\11\Mprime/32,&3l=2\Mprime,\\5\Mprime/16,&\text{otherwise}.\end{cases}
\end{aligned}
\end{equation}
In every case $\bigl|\sum_{j=1}^{M'}f_j(l)^{3}\bigr|\le\Mprime/4$, $\sum_{j=1}^{M'}f_j(l)^{4}\le3\Mprime/8$, and $\sum_{j=1}^{M'}f_j(l)^{6}\le11\Mprime/32$.

We can calculate $\braket{\Ztot^{2}}$ by combinatorics. Each site belongs to $\binom{\Mprime-1}{n-1}$ of the sets $T$ and each pair of sites to $\binom{\Mprime-2}{n-2}$ of them; with $0=(\sum_{j=1}^{M'}f_j)^{2}=\sum_{j=1}^{M'}f_j^{2}+2\sum_{i<j}f_if_j$, Eq.~\eqref{eq:config} gives
\begin{equation}\label{eq:norms}
\begin{aligned}
  \braket{\Ztot^{2}}
  &=\frac{4}{\binom{\Mprime}{n}}\Bigl[\binom{\Mprime-1}{n-1}\sum_{j=1}^{M'}f_j(l)^{2}+2\binom{\Mprime-2}{n-2}\sum_{i<j}f_i(l)f_j(l)\Bigr]
  =4\Bigl[\frac{n}{\Mprime}-\frac{n(n-1)}{\Mprime(\Mprime-1)}\Bigr]\frac{\Mprime}{2}\\
  &=\frac{2n(\Mprime-n)}{\Mprime-1}\ge n\qquad(1\le n\le\Mprime/2).
\end{aligned}
\end{equation}
The zero-sum coefficients give a root-mean-square value of $Z(l)$ of order $\sqrt{n}$ in the Dicke state, substantially smaller than the upper bound $2n$ on its operator norm restricted to the weight-$n$ sector. This difference in scaling is what our rescaling exploits.

Unlike the second moment, the fourth and sixth moments involve contributions from multiple combinations of distinct indices. We bound these moments using the following lemma.

\noindent\textbf{Lemma SA.1. (Special case of Theorem 4 in Ref.~\cite{Hoeffding1963})}\ \emph{For even $k$, $n\le\Mprime$, and real $f_1,\dots,f_{\Mprime}$, the following inequality holds:}
\begin{equation}\label{eq:convex}
  \frac{1}{\binom{\Mprime}{n}}\sum_{|T|=n}\Bigl(\sum_{j\in T}f_j\Bigr)^{k}
  \le\frac{1}{\Mprime^{n}}\sum_{j_1,\dots,j_n=1}^{\Mprime}\bigl(f_{j_1}+\dots+f_{j_n}\bigr)^{k}.
\end{equation}
The proof of Lemma SA.1 is given in Sec.~A.3.

On the right-hand side of Eq.~\eqref{eq:convex}, the indices of the summation are unrestricted. After expansion, every term in which some index occurs only once vanishes by $\sum_jf_j=0$. For $k=4$ the surviving terms are the $n$ terms with four equal indices and the $3n(n-1)$ terms with two equal pairs; for $k=6$ they are the $n$ terms with six equal indices, the $15n(n-1)$ terms with a quadruple and a pair, the $10n(n-1)$ terms with two triples, and the $15n(n-1)(n-2)$ terms with three pairs. With the power sums~\eqref{eq:powersums}, Eqs.~\eqref{eq:config} and \eqref{eq:convex} give
\begin{align}
  \braket{\Ztot^{4}}
  &\le\frac{16n}{\Mprime}\sum_{j=1}^{M'}f_j^{4}+\frac{48n(n-1)}{\Mprime^{2}}\Bigl(\sum_{j=1}^{M'}f_j^{2}\Bigr)^{2}
  \le6n+12n(n-1)\le12n^{2},\label{eq:z4}\\
  \braket{\Ztot^{6}}
  &\le\frac{64n}{\Mprime}\sum_{j=1}^{M'}f_j^{6}+\frac{960n(n-1)}{\Mprime^{2}}\sum_{j=1}^{M'}f_j^{4}\sum_{j=1}^{M'}f_j^{2}
  +\frac{640n(n-1)}{\Mprime^{2}}\Bigl(\sum_{j=1}^{M'}f_j^{3}\Bigr)^{2}
  +\frac{960n(n-1)(n-2)}{\Mprime^{3}}\Bigl(\sum_{j=1}^{M'}f_j^{2}\Bigr)^{3}\nonumber\\
  &\le22n+180n(n-1)+40n(n-1)+120n(n-1)(n-2)=120n^{3}-140n^{2}+42n\le120n^{3}.\label{eq:z6}
\end{align}

These moments bound the infidelity and the success probability. Let $u\equiv(\sin[\tau\Ztot]-\tau\Ztot)\Dicke{n}{\Mprime}$. Since $|\sin x-x|\le|x|^{3}/6$ for all real $x$, Eq.~\eqref{eq:z6} gives $\|u\|^{2}\le\tau^{6}\braket{\Ztot^{6}}/36\le\tfrac{10}{3}\tau^{6}n^{3}$. For nonzero vectors $u_1$ and $u_2$, the fidelity $F$ of their normalizations obeys $1-F\le\|u_1-u_2\|^{2}/\|u_1\|^{2}$, since $\|u_1\|^{2}(1-F)$ is the squared distance from $u_1$ to the line spanned by $u_2$. With $u_1=\tau\Ztot\Dicke{n}{\Mprime}$ and $u_2=\sin[\tau\Ztot]\Dicke{n}{\Mprime}$, the normalized vectors are $\ket A$ and $\ket{\tilde A}$ of Eq.~\eqref{eq:ASdef}, $u_2-u_1=u$, and $\|u_1\|^{2}=\tau^{2}\braket{\Ztot^{2}}\ge\tau^{2}n$. Hence,
\begin{equation}\label{eq:infbound}
  1-F\le\frac{\|u\|^{2}}{\|u_1\|^{2}}\le\frac{\tfrac{10}{3}\tau^{6}n^{3}}{\tau^{2}n}=\frac{10}{3}\tau^{4}n^{2}.
\end{equation}
Any $0<\tau\le(3/10)^{1/4}\varepsilon^{1/4}/\sqrt n$ therefore satisfies Eq.~\eqref{eq:target}.
For the success probability $p_{\mathrm{succ}}=\|\sin[\tau Z(l)]|D^{M'}_n\rangle \|^2$, $\sin^{2}x\ge x^{2}-x^{4}/3$ gives
\begin{equation}\label{eq:pbound}
  p_{\mathrm{succ}}\ge\tau^{2}\braket{\Ztot^{2}}-\frac{\tau^{4}}{3}\braket{\Ztot^{4}}
  \ge\tau^{2}n-4\tau^{4}n^{2}.
\end{equation}
Both bounds depend on $\tau$ only through $\tau\sqrt n$. Fixing $\tau=\varepsilon^{1/4}/(2\sqrt n)$ makes $\tau^{4}n^{2}=\varepsilon/16$, so that
\begin{equation}\label{eq:epsguarantee}
  1-F\le\frac{5}{24}\varepsilon\le\varepsilon,
  \qquad
  p_{\mathrm{succ}}\ge\frac{\sqrt\varepsilon}{4}-\frac{\varepsilon}{4}\ge\frac{\sqrt\varepsilon}{16}
  \qquad\left(\varepsilon\le\tfrac{1}{2}\right).
\end{equation}
This gives a positive lower bound on the success probability that depends on $\varepsilon$, independently of $n$ and the system size; hence, $O(1)$ amplification rounds suffice. Choosing $\tau =1 /(2n)$ based on the operator-norm bound instead yields $p_{\mathrm{succ}} = \Theta (1/n)$ and $\Theta (\sqrt{\Mprime})$ rounds at half filling.

We next evaluate the success probability and fidelity more precisely. For $\Ztot\ket{T}=\lambda_T\ket{T}$, where $\lambda_T=-2\sum_{j\in T}f_j(l)$, let $\mathbb{E}_T$ denote the uniform average over all subsets $T$ with $|T|=n$. Since $\Dicke{n}{\Mprime}$ is the equal superposition of these basis states, we obtain
\begin{equation}\label{eq:avg_lambda}
  p_{\mathrm{succ}} =\mathbb{E}_T[\sin^2(\tau\lambda_T)],
  \qquad
  F=\frac{\bigl(\mathbb{E}_T[\lambda_T\sin(\tau\lambda_T)]\bigr)^2}{\mathbb{E}_T[\lambda_T^2]\mathbb{E}_T[\sin^2(\tau\lambda_T)]
  }.
\end{equation}
Defining $x_T=\lambda_T/\braket{\Ztot^2}^{1/2}$ and $s=\tau\braket{\Ztot^2}^{1/2}$, we have $\mathbb{E}_T[x_T^2]=1$ and
\begin{equation}\label{eq:avg_x}
  p_{\mathrm{succ}}=\mathbb{E}_T[\sin^2(sx_T)], \qquad
  F=\frac{\bigl(\mathbb{E}_T[x_T\sin(sx_T)]\bigr)^2}{\mathbb{E}_T[\sin^2(sx_T)]}.
\end{equation}
The eigenvalues $\lambda_T$ correspond to sums of $n$ values sampled without replacement from $\{-2f_j(l)\}_{j=1}^{\Mprime}$. In the limit $\Mprime\to\infty$ with $n/\Mprime$ fixed and nonzero, the distribution of $x_T$ approaches the standard normal distribution $e^{-x^2/2}/\sqrt{2\pi}$. For fixed $s$, the averages in Eq.~\eqref{eq:avg_x} therefore approach
\begin{align*}
  \mathbb{E}_T[\sin^2(sx_T)] &\longrightarrow \frac{1}{2\sqrt{2\pi}} \int_{-\infty}^{\infty} e^{-x^2/2}[1-\cos(2sx)]\,dx =\frac12(1-e^{-2s^2}),\\
  \mathbb{E}_T[x_T\sin(sx_T)] &\longrightarrow -\frac{d}{ds}
  \left[
    \frac{1}{\sqrt{2\pi}}
    \int_{-\infty}^{\infty}
    e^{-x^2/2}\cos(sx)\,dx
  \right]
  =s\,e^{-s^2/2},
\end{align*}
where we used $\int_{-\infty}^{\infty}e^{-x^2/2}\cos(tx)\,dx =\sqrt{2\pi}e^{-t^2/2}$. Substituting these results into Eq.~\eqref{eq:avg_x} gives, in this limit,
\begin{equation}\label{eq:gauss_tradeoff}
  \displaystyle p_{\mathrm{succ}}(s)=\frac{1}{2}\left(1-e^{-2s^{2}}\right),\qquad F(s)=\frac{2s^{2}e^{-s^{2}}}{1-e^{-2s^{2}}}.
\end{equation}
For small $s$, these expressions yield $1-F\simeq s^{4}/6$ and $p_{\mathrm{succ}}\simeq\sqrt{6(1-F)}$. Exact evaluation of $p_{\mathrm{succ}}$ and $F$ from the $\Ztot$ spectrum agrees with Eq.~\eqref{eq:gauss_tradeoff} to within $10^{-3}$ at $s=0.495$ already at $\Mprime=64$ at half filling (for $n=1$ and fixed $l$, the large-$M'$ limit is instead an arcsine distribution). The choice $s=0.495$ in Eq.~(\mainref{eq:tau_choice}{12}) is motivated by the Gaussian-limit expressions (\ref{eq:gauss_tradeoff}). In the finite-size cases examined numerically ($M'=4, 6, 8, 10, 12, l=1,$ and $n=1,2$, and $M'/2$), it yields $1-F\leq 0.01$ and $0.19\leq p_{\mathrm{succ}}\leq0.23$, for which a single amplification round gives a success probability above $0.95$. Rigorous uniform fidelity and success-probability guarantees for a more conservative choice of $\tau$ are given in Eq.~(\ref{eq:epsguarantee}).

A uniform positive lower bound on the success probability also follows for the choice (\mainref{eq:tau_choice}{12}). Using Eqs.~(\ref{eq:norms}), (\ref{eq:z4}), and (\ref{eq:pbound}), we obtain
\begin{align}
p_{\mathrm{succ}}
\geq
s^2-\frac{s^4\langle Z(l)^4\rangle}{3\langle Z(l)^2\rangle^2}
\geq s^2-4s^4>0
\qquad (s=0.495).\label{eq:inequality_pcuss}
\end{align}
Thus, $K=O(1)$ holds uniformly in $n$ and $M'$ for this choice as well.

\begin{figure}[htbp]
\centering
\makebox[0.835\textwidth][s]{\textbf{(a)} $U_{\mathrm{BE}}$\hfill\textbf{(b)} $R_0$}\\[2pt]
\circuitpdf[0.48549]{\includegraphics[hiresbb,width=0.72\textwidth]{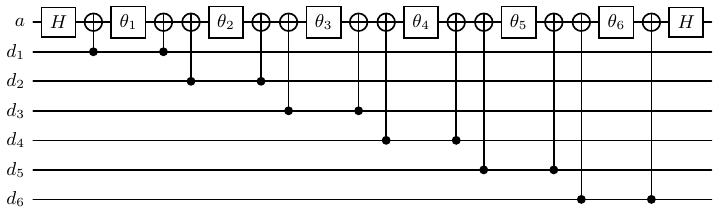}}\hspace{0.04\textwidth}%
\circuitpdf[0.48552]{\includegraphics[hiresbb,width=0.075\textwidth]{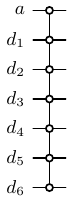}}
\caption{
Explicit circuits for the block-encoding and reflection of the Grover iterate $\mathcal G$ [Fig.~\mainref{fig:circuit}{1}(b) in the main text] at $(\Mprime,n,l)=(6,2,1)$; the third module, Dicke preparation $U_{\mathrm D}$, is the B\"artschi--Eidenbenz protocol~\cite{BartschiEidenbenz2019}. (a)~Block encoding $U_{\mathrm{BE}}$: Hadamards on the ancilla $a$ sandwich the six commuting two-qubit phases $e^{i\theta_j Z_{\mathrm{a}} Z_j}$ with $\theta_j=\tau f_j(1)$ and $\tau$ given by Eq.~(\mainref{eq:tau_choice}{12}); each phase is compiled as $\mathrm{CNOT}\cdot e^{i\theta_j Z_{\mathrm{a}}}\cdot\mathrm{CNOT}$, where the box labeled by $\theta_j$ denotes $e^{i\theta_j Z_{\rm a}}$. (b)~The reflection $R_0$ on ancilla plus data, realized up to a global phase by the $(\Mprime+1)$-qubit controlled-$Z$ with all controls open (open circles denote control on $\ket{0}$), i.e., by the phase $-1$ on $\ket{0\cdots0}$ alone. The flag-$\ket{1}_{\mathrm{a}}$ branch of $U_{\mathrm{BE}}$ equals $i\sin[\tau\Ztot]$ on arbitrary inputs.
}
\label{fig:components}
\end{figure}

Figure~\ref{fig:components} shows the circuits for the block-encoding unitary $U_{\rm BE}$ and the reflection $R_0$ used in the Grover iterate for $(\Mprime,n,l)=(6,2,1)$. All three modules (these two and the Dicke preparation $U_{\mathrm D}$) act on the effective chain only and are therefore identical for the DWC and Onsager scar models (and for every model whose scar tower maps onto the Dicke tower): the model enters solely through $\Phi^{-1}$ and $\Mprime$ (Fig.~\ref{fig:phi_circuit}).

\subsection*{A.2\quad Oracle depth and the cost of fan-out}

Here, we discuss the details of oracle depth and the cost of fan-out. The following depth estimates assume all-to-all connectivity and $\Theta (M')$ clean ancillas, reused by both tree constructions below. The block-encoding oracle factorizes into commuting two-qubit rotations, $e^{i\tau Z_{\mathrm{a}}\otimes\Ztot}=\prod_{j=1}^{M'}e^{i\tau f_j(l) Z_{\mathrm{a}} Z_j}$, but all $\Mprime$ of them couple to the single ancilla $a$; commuting gates that share a qubit cannot be applied simultaneously. Therefore, the naive depth is $O(\Mprime)$. A coherent CNOT fan-out tree distributes the ancilla's computational-basis control across $\Mprime$ auxiliary qubits in depth $O(\log M')$. The rotations $e^{i\tau f_j(l) Z_j Z_{\mathrm{a}_j}}$ can then be applied in parallel, after which the fan-out is uncomputed. The resulting oracle depth is $O(\log\Mprime)$.

A similar tree construction implements the reflection $R_0=2|0\cdots0\rangle\langle0\cdots0|-I$ in $O(\log M')$ depth. Up to a global phase and conjugation by single-qubit $X$ gates, it is the $(M'+1)$-qubit controlled-$Z$ gate. A balanced binary tree of Toffoli gates computes the logical AND of the qubits into $O(\Mprime)$ clean ancillas, a controlled-$Z$ imprints the phase at the root, and the tree is uncomputed. This construction has depth $O(\log\Mprime)$~\cite{Barenco1995}. For the gate-noise simulations in Sec.~C, we instead use the one-clean-ancilla decomposition of Ref.~\cite{KhattarGidney2025}.

\label{subsec:proof_of_lemmaSA1}\subsection*{A.3\quad Proof of Lemma SA.1}

Let us consider $n$ independent draws with replacement, each uniformly distributed over $\{1,\ldots,\Mprime\}$. Let $N_j$ denote the number of times index $j$ is selected, and define the random variable
\begin{equation}
    S=\sum_{j=1}^{\Mprime}N_jf_j,\qquad\sum_{j=1}^{\Mprime}N_j=n.
\end{equation}
Here, $\mathbb E[\cdot]$ denotes expectation over the original $n$ independent uniform draws. Since each index is selected with probability $1/\Mprime$ in each draw, we have $\mathbb E[N_j]=n/\Mprime$. Thus, by linearity of expectation,
\begin{equation}
    \mathbb E[S]=\sum_{j=1}^{\Mprime}f_j\mathbb E[N_j]=\frac{n}{\Mprime}\sum_{j=1}^{\Mprime}f_j.
\end{equation}

Next, construct a subset $T$ by collecting all distinct indices from the original draws and adding uniformly chosen indices not already selected until $|T|=n$. The original draws treat all index labels equally, and the additional indices are chosen uniformly from those not already included. Thus, the entire construction is invariant under permutations of the index labels, and all subsets of size $n$ have the same probability. Since there are $\binom{\Mprime}{n}$ such subsets, we obtain
\begin{equation}
    \Pr(T=t)=\frac{1}{\binom{\Mprime}{n}}, \qquad |t|=n.
\end{equation}

We now evaluate the conditional expectation $\mathbb E[S | T]$. Here, $\mathbb E[S | T=t]$ denotes the average of $S$ over the original draws and the subsequent uniform selection of additional indices, conditioned on the resulting subset being $T=t$. This average is weighted by the corresponding conditional probabilities. For a fixed subset $T=t$, every index selected in the original draws belongs to $t$. Consequently, $N_j=0$ for $j\notin t$ and $\sum_{j\in t}N_j=n$. Furthermore, permutations of the indices within $t$ leave both the sampling procedure and the condition $T=t$ unchanged. Therefore, the conditional distribution of the multiplicities $N_j$ is invariant under such permutations. Hence,
\begin{equation}
    \mathbb E[N_j | T=t]=c_t, \qquad j\in t,
\end{equation}
where $c_t$ is independent of $j$. Taking the conditional expectation of $\sum_{j\in t}N_j=n$ gives
\begin{equation}
    \sum_{j\in t}\mathbb E[N_j | T=t] =nc_t=n.
\end{equation}
Thus, $c_t=1$, and
\begin{equation}
    \mathbb E[N_j | T=t]=
    \begin{cases}
        1, & j\in t,\\
        0, & j\notin t.
    \end{cases}
\end{equation}
Using the linearity of conditional expectation, we obtain
\begin{align}
    \mathbb E[S | T]
    &=\mathbb E\left[
        \sum_{j=1}^{\Mprime}N_jf_j
        \,\middle|\,T\right]\notag\\
    &=\sum_{j=1}^{\Mprime}
        f_j\mathbb E[N_j | T]\notag\\
    &=\sum_{j\in T}f_j.
\end{align}

Let $\mathbb E_T[\cdot]$ denote the uniform average over all subsets $T$ with $|T|=n$. Using the above result and $\Pr(T=t)=1/\binom{\Mprime}{n}$, the left-hand side of Eq.~\eqref{eq:convex} can be expressed as $\mathbb E_T[(\mathbb E[S | T])^k]$. Since $x\mapsto x^k$ is convex for even $k$, the conditional Jensen inequality and the law of total expectation~\cite{Williams1991} give
\begin{align}
    \frac{1}{\binom{\Mprime}{n}}\sum_{|T|=n}\left(\sum_{j\in T}f_j\right)^k
    &=\mathbb E_T\left[
        (\mathbb E[S | T])^k
    \right]\notag\\
    &\leq\mathbb E_T\left[
        \mathbb E[S^k | T]
    \right]\notag\\
    &=\sum_{|t|=n}\Pr(T=t)
        \mathbb E[S^k | T=t]\notag\\
    &=\mathbb E[S^k]\notag\\
    &=\frac{1}{\Mprime^n}
    \sum_{j_1,\ldots,j_n=1}^{\Mprime}
    (f_{j_1}+\cdots+f_{j_n})^k.
\end{align}
The inequality follows from the conditional Jensen inequality, while the sum over $t$ equals $\mathbb E[S^k]$ by the law of total expectation. The last equality follows because each sequence $(j_1,\ldots,j_n)$ of the original independent uniform draws occurs with probability $1/(\Mprime)^n$.

\section*{B.\quad Domain-wall-conserving and Onsager scar models}

\subsection*{B.1\quad Models and towers}

The DWC and Onsager scar Hamiltonians and their ladder operators are defined in Eqs.~(\mainref{eq:hdwc}{15})--(\mainref{eq:qons}{19}) of the main text, following Eqs.~(83)--(88) and (94)--(104) of Ref.~\cite{Kunimi2025}. The boundary conventions reduce the Onsager edge terms to the two-site operators $ h_{{\mathrm{c}},1}=\half P'_1 P_2$ and $ h_{{\mathrm{c}},M}=\half P_{M-1} P'_M$. We use $h/J=0.7$ and $J_z/J=1.3$ for the DWC model and $J_{\rm c}/J=1$ and $h/J=0.7$ for the Onsager scar model.

The DWC tower $\ket{S_n}\propto( Q^\dagger_{\mathrm{DWC}})^n \ket{0}^{\otimes M}$ consists of $n$ mutually nonadjacent excitations carrying a sublattice sign, with energies $E_n=h(M/2-n)+J_z\bigl[(M+1)/4-n\bigr]$; the effective chain has $\Mprime=M-n+1$ sites [$\mathcal N_{\mathrm{DWC}}(M,n)=\binom{M-n+1}{n}$ configurations].

The Onsager tower $\ket{S_n}\propto( Q^\dagger_{\mathrm{Ons}})^n\ket{0}^{\otimes M}$ consists of even-length clusters of adjacent up spins ($n$ pairs in total), with energies $E_n=h(M/2-2n)$; the effective chain has $\Mprime=M-n$ sites [$\mathcal N_{\mathrm{Ons}}(M,n)=\binom{M-n}{n}$ configurations].

\subsection*{B.2\quad Explicit form of $\Phi^{-1}$}

Here, we show the explicit form of the isometry $\Phi^{-1}$. The action of the three primitives on basis configurations is given in the main text. The staggered phase quoted there is the Onsager form; the DWC form, $(-1)^{\sum_k(q_k+k+1)}$, differs from it only by the global phase $(-1)^{n}$.

The coefficients $f_j(l)$ are functions of the effective coordinate, not of the physical position. The operator $\Ztot$ creates the configuration amplitudes $\lambda_T=-2\sum_{j\in T}f_j(l)$ on the effective chain, and $\Phi^{-1}$ carries them unchanged to a single signed physical configuration, assigning no amplitude structure to the appended qubits.  The earlier construction of Ref.~\cite{Kunimi2024} uses a different modulation in the physical coordinates. The parent-Hamiltonian construction can reduce the energy variance~\cite{Kunimi2025}. The mapping is exact for both models: it preserves the allowed configurations and staggered phases and is injective on the effective-chain Hilbert space.

\begin{figure}[htbp]
\centering
\textbf{(a)}\\[2pt]
\circuitpdf[0.51953]{\includegraphics[hiresbb,width=0.55\textwidth]{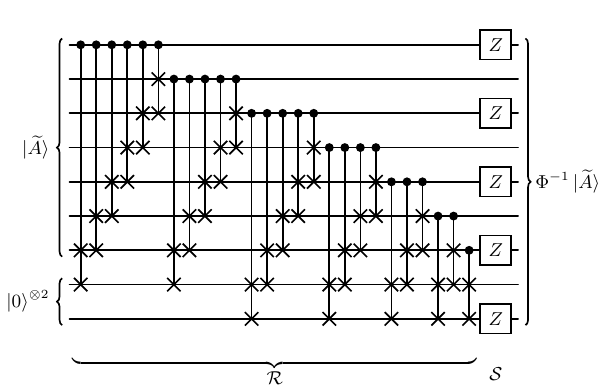}}\\[5pt]
\textbf{(b)}\\[2pt]
\circuitpdf[0.51666]{\includegraphics[hiresbb,width=0.92\textwidth]{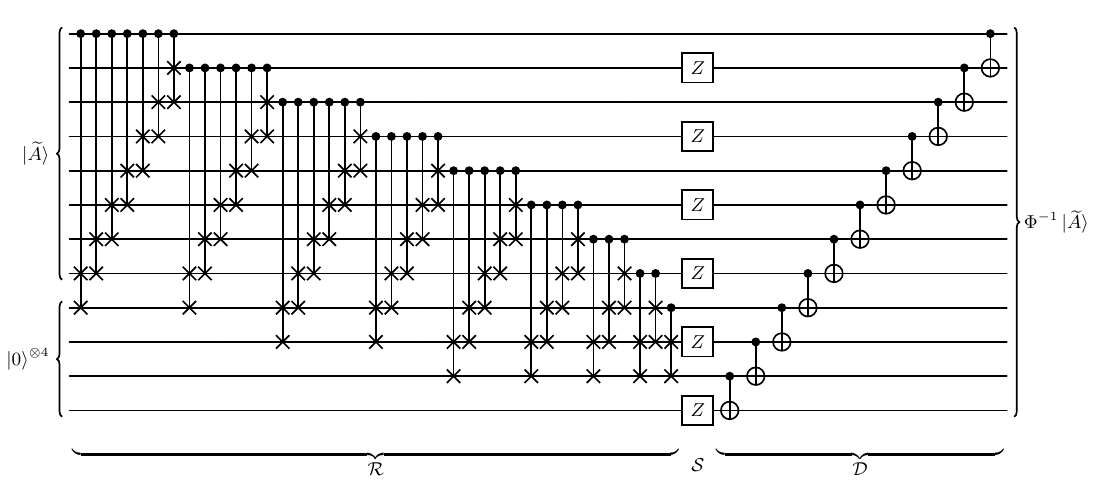}}
\caption{
Examples of reversible circuits implementing $\Phi^{-1}=\mathcal{D}\circ \mathcal{S}\circ \mathcal{R}$ for two models with different values of $\Mprime$ and $n$: (a) the DWC model at $(\Mprime,n)=(7,3)$ [$M=\Mprime+n-1=9$ physical sites] and (b) the Onsager scar model at $(\Mprime,n)=(8,4)$ [$M=\Mprime+n=12$]. The braces indicate the three circuit components. The top wires carry the postselected state $|\widetilde{A}\rangle$ defined in Eq.~(\ref{eq:ASdef}), which is mapped to the physical state $\Phi^{-1}|\widetilde{A}\rangle$. The appended qubits are initialized in $|0\rangle$, and no ancillas are used. With a Dicke-state input $|D_n^{M'}\rangle$, the same circuits prepare the corresponding exact scar states.
}
\label{fig:phi_circuit}
\end{figure}

Figure~\ref{fig:phi_circuit} shows the complete reversible circuits for the DWC and Onsager scar models. The operation $\mathcal{D}$ is an ancilla-free sweep of $M-1$ CNOTs, $\mathcal{S}$ a single diagonal phase layer, and $\mathcal{R}$ an ancilla-free sweep of $\Theta(M^{2})$ Fredkin gates, which we call the \emph{conveyor}.  The gate sequence depends only on $M$ and can therefore be used for any scar index $n$, with the dependence on $n$ entering through $\Mprime$. For the noise simulations, we omit gates acting trivially on all reachable input states, reducing the gate count without changing its asymptotic scaling.

Two special cases admit routing cost $\Theta(M)$ in both gate count and depth. At half filling in the DWC model, the effective chain contains a single hole, and the routing circuit reduces to the \emph{hole ladder}: a cascade that accumulates hole flags, one layer of pairwise disjoint Fredkin gates, and an exact flag erasure. At $n=2$, only the rightmost excitation is shifted by one lattice site, and one carry ancilla suffices; we call this circuit the \emph{particle ladder}. For intermediate scar indices ($3\leq n\leq \Mprime-2$), where several excitations move by different distances, we have not identified a more efficient implementation than the conveyor. We therefore use the conveyor circuit in our gate-noise simulations.

\subsection*{B.3\quad Relaxation dynamics of the DWC and Onsager scar models}

\begin{figure}[htbp]
\centering
\includegraphics[width=\linewidth]{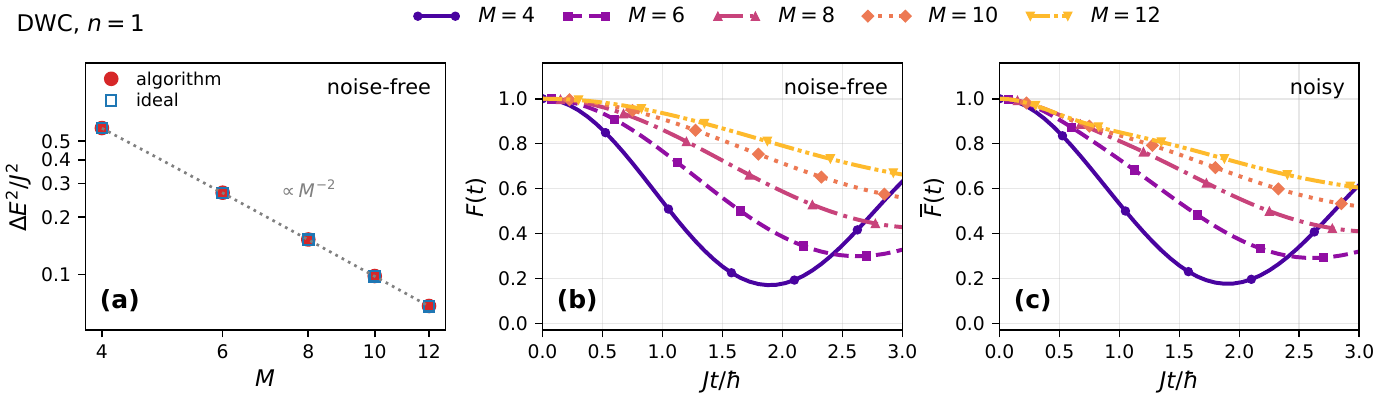}\\[0.9em]
\includegraphics[width=\linewidth]{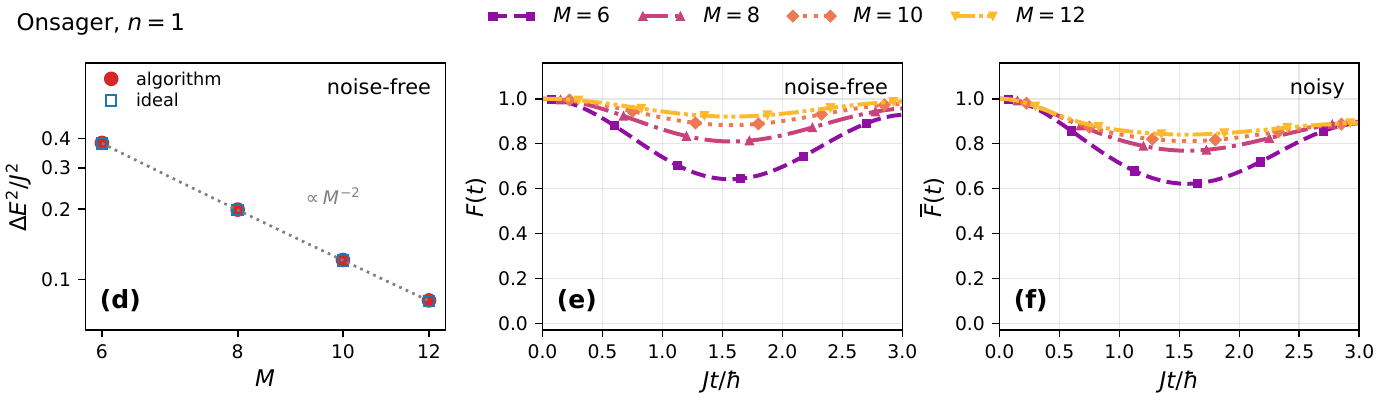}
\caption{Numerical results for the DWC and Onsager scar models for $n=1$ and $l=1$. \emph{Top:} DWC chain, sizes $M=4,6,8,10$, and $12$. \emph{Bottom:} Onsager scar model, sizes $M=6,8,10$, and $12$. Left (a,d): Noise-free energy variance $\Delta E^2/J^2$ versus $M$ for the algorithm output and the ideal AQMBS state, both scaling as $M^{-2}$. Middle (b,e): Noise-free self-overlap fidelity $F(t)=|\braket{\psi (0) | \psi (t)}|^2$. Right (c,f): Noise-averaged fidelity $\overline{F}(t)$ under depolarizing gate noise with error rates $(\varepsilon_1, \varepsilon_2) = (10^{-4}, 1.5 \times 10^{-3})$. }
\label{fig:sm_constrained_n1}
\end{figure}

\begin{figure}[htbp]
\centering
\includegraphics[width=0.72\linewidth]{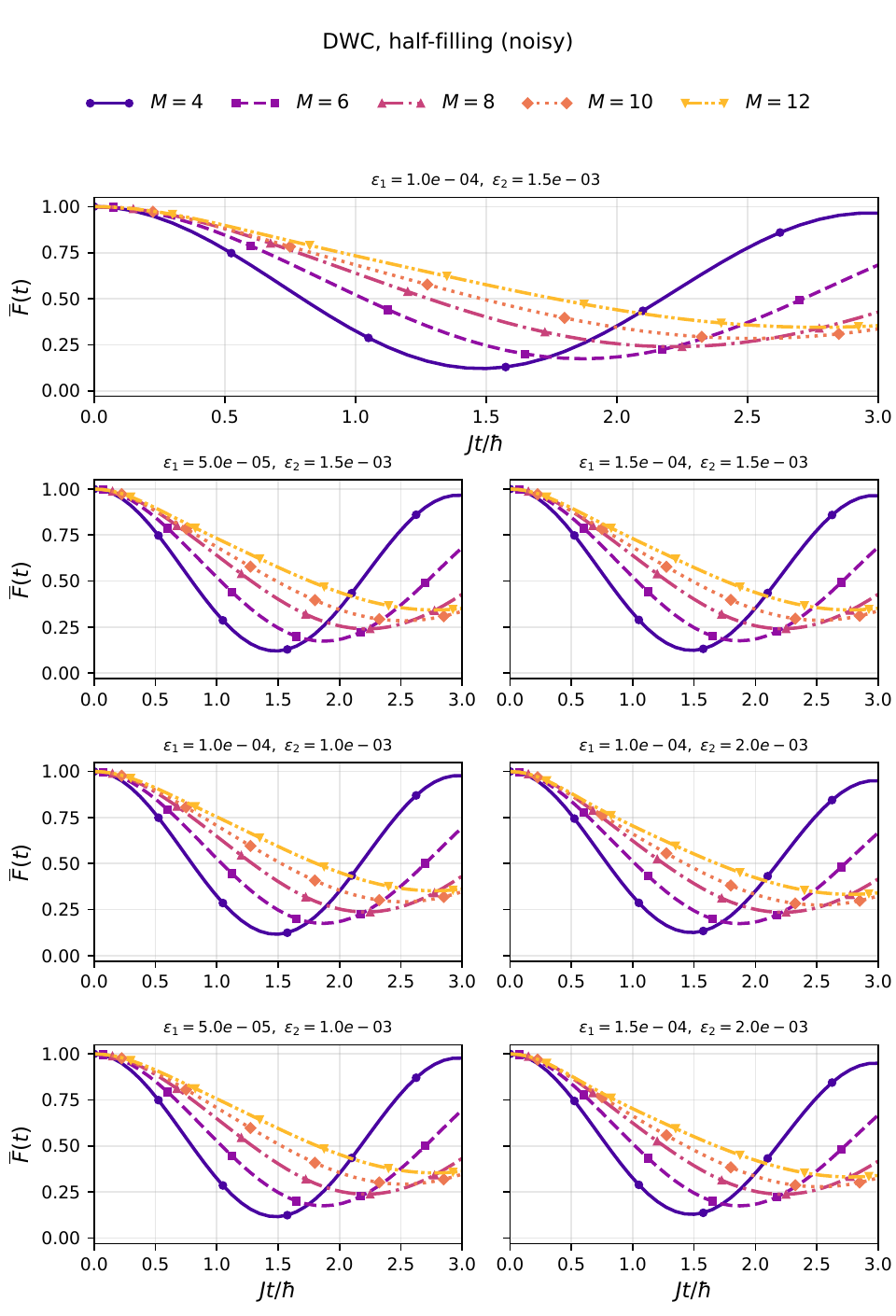}
\caption{DWC model at half filling ($n=M/2$) under depolarizing gate noise for $l=1$. Each panel corresponds to a distinct pair of one- and two-qubit error rates $(\varepsilon_1,\varepsilon_2)$, printed above it: the top panel is the central pair $(10^{-4},1.5\times10^{-3})$; the pairs below vary the one-qubit rate $\varepsilon_1$ (row~1), the two-qubit rate $\varepsilon_2$ (row~2), and both together (row~3) by one grid step in each direction.}
\label{fig:sm_dwc_half}
\end{figure}

\begin{figure}[htbp]
\centering
\includegraphics[width=0.72\linewidth]{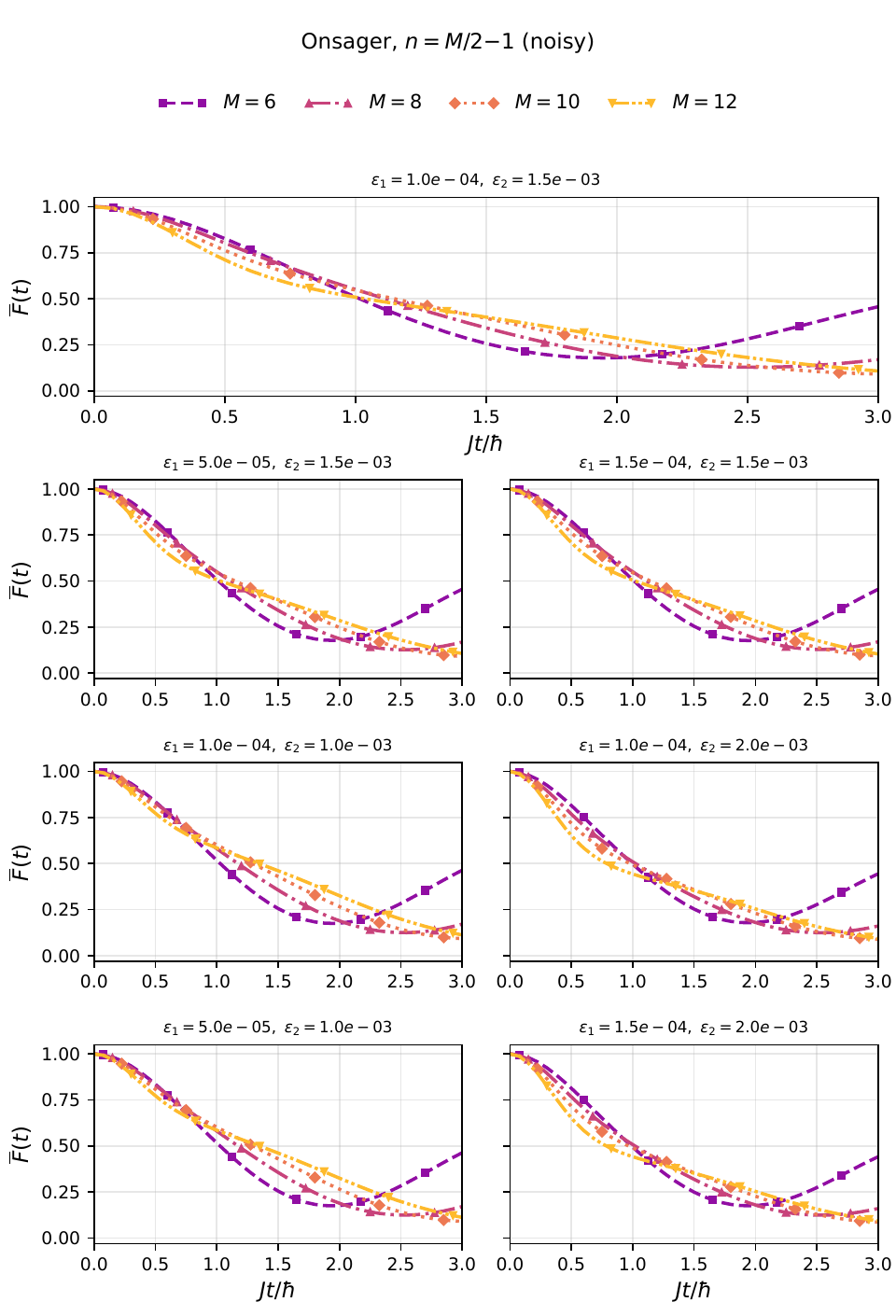}
\caption{Onsager scar model near the maximal scar index $n=M/2-1$ under depolarizing gate noise, sizes $M=6,8,10,12$ for $l=1$. Panels as in Fig.~\ref{fig:sm_dwc_half}.}
\label{fig:sm_onsager_half}
\end{figure}

Figures~\ref{fig:sm_constrained_n1}--\ref{fig:sm_onsager_half} extend the numerical analysis to the DWC and Onsager scar models. Each algorithm-prepared state is mapped to the physical Hilbert space by $\Phi^{-1}$ and evolved under the corresponding Hamiltonian. As in the spin-1 {\it XY} model (Fig.~\mainref{fig:relax}{2} in the main text), the self-overlap fidelity decays more slowly with increasing $M$ [Figs.~\ref{fig:sm_constrained_n1}(b) and \ref{fig:sm_constrained_n1}(e)]. The energy variances of the algorithm-prepared and ideal AQMBS states scale as $M^{-2}$ and agree to within 2\% [Figs.~\ref{fig:sm_constrained_n1}(a) and \ref{fig:sm_constrained_n1}(d)].

At the central error rates, the initial decay of $\overline{F}(t)$ is generally slower for larger $M$ at $n=1$ [Figs.~\ref{fig:sm_constrained_n1}(c) and \ref{fig:sm_constrained_n1}(f)]. The DWC model shows the same trend at half filling (Fig.~\ref{fig:sm_dwc_half}). For the Onsager scar model at $n=M/2-1$, however, the $M=12$ curve lies below those for $M=6$--$10$ at short times (Fig.~\ref{fig:sm_onsager_half}). The same initial reversal occurs in the spin-1 {\it XY} model at half filling. The circuits in the two cases exhibiting this reversal use $\Theta(M^2)$ gates, whereas the DWC preparation at half filling uses $\Theta(M)$ gates (Secs.~B.2 and C).

For the Onsager scar model, we omit $M=4$. At $n=1$, the corresponding AQMBS state has support on only two energy eigenstates, leading to a two-level revival specific to this small system. This finite-size effect is absent for $M\geq6$ and should not be interpreted as a characteristic signature of AQMBS dynamics.

\clearpage
\section*{C.\quad Noise robustness: gate error rates and amplification depth}\label{app:noise_robustness}

To assess how gate error rates and amplification depth affect the success probability and prepared-state fidelity, we simulate trajectories of the complete circuit---Dicke preparation, the block encoding, $K$ rounds of amplitude amplification, and the routing isometry $\Phi^{-1}$---with depolarizing noise after every gate. Denoting by $\rho$ the density matrix of all qubits right after a gate, we apply the depolarizing channel of rate $\varepsilon_1$ after each one-qubit gate and the two-qubit depolarizing channel of rate $\varepsilon_2$ after each two-qubit gate~\cite{Fowler2012},
\begin{align}
\label{eq:depol1}
\mathcal{E}_1 ^{(j)}(\rho)&=(1-\varepsilon_1)\,\rho+\frac{\varepsilon_1}{3}\sum_{P\in\{X,Y,Z\}}P_j\rho P_j,\\
\label{eq:depol2}
\mathcal{E}_2 ^{(j,k)}(\rho)&=(1-\varepsilon_2)\,\rho+\frac{\varepsilon_2}{15}\sum_{\substack{P,Q\in\{I,X,Y,Z\}\\(P,Q)\neq(I,I)}}(P_j Q_k)\,\rho\,(P_j Q_k),
\end{align}
where $P_j$ denotes the operator $P\in \{I,X,Y,Z\}$ acting on qubit $j$ and as the identity on all other qubits, with $I$, $X$, $Y$, and $Z$ being the single-qubit identity and Pauli operators, respectively. For the time-evolution and entanglement-entropy calculations, we prepare the initial states without amplification ($K=0$). For each data point in the noise simulations, we compute 2400 trajectories before postselection. The routing layer is executed gate by gate at the given $(M,n)$: at $n=1$ the route $\mathcal{R}$ is empty (the conveyor acts as the identity on the reachable sector), at $n=2$ it is the particle ladder of Sec.~B.2, at DWC half filling the hole ladder, and otherwise the conveyor. In our noise model, each Fredkin gate contributes eight two-qubit error locations, while each Toffoli gate contributes six. The number of error locations associated with $\mathcal{R}$ is therefore zero at $n=1$, $\Theta(M)$ for either ladder, and $\Theta(M^2)$ for the conveyor. The auxiliary qubits used in the ladder circuits return deterministically to $\ket{0}$ in the noiseless limit. Under noise, their measurement outcomes are sampled according to the Born rule in each trajectory. For $K\ge1$, we implement $R_0$ using the one-clean-ancilla decomposition of a multi-controlled $X$ gate proposed by Khattar and Gidney~\cite{KhattarGidney2025}. In our compilation, each application of $R_0$ requires $6M'-6$ CNOT gates and single-qubit gates. 

At half filling in the DWC model, $n=\Mprime-1$, the Dicke state $\Dicke{\Mprime-1}{\Mprime}$ can be prepared with $\Theta(M)$ gates by applying a global bit flip to the single-excitation Dicke state $\Dicke{1}{\Mprime}$. Together with the $\Theta(M)$-gate hole-ladder routing, this gives a $\Theta(M)$-gate preparation circuit. Table~\ref{tab:hw} lists representative gate errors reported for current quantum hardware, using the error metric adopted in each reference. The trapped-ion and superconducting values are average gate infidelities $r$, i.e., one minus the gate fidelity averaged uniformly over pure input states. For the channels in Eqs.~\eqref{eq:depol1} and~\eqref{eq:depol2}, the parameters $\varepsilon_1$ and $\varepsilon_2$ represent the total probabilities of nonidentity Pauli errors. The corresponding average gate infidelities are $r_1=2\varepsilon_1/3$ and $r_2=4\varepsilon_2/5$~\cite{Nielsen2002}. For the neutral-atom values, we treat the reported gate errors as effective average gate infidelities when determining the depolarizing-channel parameters. The channel parameters therefore follow from the reported values as
\begin{equation}
\varepsilon_1=\frac{3}{2}\,r_1,\qquad \varepsilon_2=\frac{5}{4}\,r_2 .
\label{eq:eps_from_r}
\end{equation}
We use the values in Table \ref{tab:hw} as a guide to select the range of channel parameters for our numerical simulations.

The near-term claim in the main text concerns $n=O(1)$ rather than $n=1$ alone. We therefore report results for intermediate values of $n$ at the central error rates and $M=12$. For $n=2,3$ at $M=12$, one amplification round gives success probabilities of $0.63$--$0.72$ at the central error rates; Table~\ref{tab:midorder} also reports the corresponding prepared-state fidelities. The prepared-state fidelity decreases more strongly with $n$, particularly in the constrained models. At $n=3$ the routing is the $\Theta(M^{2})$ conveyor rather than the $\Theta(M)$ particle ladder (Sec.~B.2). 
\clearpage
\begin{table}[htbp]
\caption{Representative single- and two-qubit gate errors reported for current quantum hardware, together with the corresponding depolarizing-channel parameters \(\varepsilon_1\) and \(\varepsilon_2\), used to guide the choice of the scanned noise range. The superconducting~\cite{IBMHeron} and trapped-ion~\cite{QuantinuumH1} values are vendor-reported, with IBM calibration data updated daily (accessed November 16, 2026). The neutral-atom values are the individually addressed single-qubit Raman-gate error reported in Ref.~\cite{Bluvstein2024} and the CZ-gate error reported in Ref.~\cite{Evered2026}, both obtained using reconfigurable neutral-atom arrays. The neutral-atom value of $\varepsilon_1$ lies above the local grid used for the time-evolution simulations in Figs.~\ref{fig:sm_dwc_half}--\ref{fig:sm_spin1xy_half}, but is included in the broader sweeps in Figs.~\ref{fig:sm_heat_spin1xy_n1}--\ref{fig:sm_heat_onsager_half}. Within the scanned range, however, the prepared-state fidelity is more sensitive to $\varepsilon_2$ than to $\varepsilon_1$ (Figs.~\ref{fig:sm_heat_spin1xy_n1}--\ref{fig:sm_heat_onsager_half}).  The time-evolution simulations use selected combinations of $\varepsilon_1\in\{0.5,1.0,1.5\}\times10^{-4}$ and $\varepsilon_2\in\{1.0,1.5,2.0\}\times10^{-3}$, with central values $(\varepsilon_1, \varepsilon_2) = (1.0 \times 10^{-4}, 1.5\times 10^{-3})$.}
\label{tab:hw}
\begin{ruledtabular}
\begin{tabular}{lccccc}
Platform & \multicolumn{2}{c}{Reported error} & \multicolumn{2}{c}{Channel parameter} & Ref. \\
\cline{2-3}\cline{4-5}
 & 1Q & 2Q & $\varepsilon_1$ & $\varepsilon_2$ & \\
\colrule
Trapped ion (Quantinuum H1-1) & $1.8\times10^{-5}$ & $9.7\times10^{-4}$ & $2.7\times10^{-5}$ & $1.2\times10^{-3}$ & \cite{QuantinuumH1} \\
Superconducting (IBM Heron r3, \texttt{ibm\_boston}) & $1.6\times10^{-4}$ & $1.2\times10^{-3}$ & $2.4\times10^{-4}$ & $1.5\times10^{-3}$ & \cite{IBMHeron} \\
Neutral atom (reconfigurable Rydberg array) & $8.8\times10^{-4}$ & $1.5\times10^{-3}$ & $1.3\times10^{-3}$ & $1.8\times10^{-3}$ & \cite{Bluvstein2024,Evered2026} \\
\end{tabular}
\end{ruledtabular}
\end{table}

\begin{table}[htbp]
\caption{Intermediate scar indices ($n=2,3$) at the central rates
$(\varepsilon_1,\varepsilon_2)=(10^{-4},1.5\times10^{-3})$ and $M=12$, after a single amplification round. $p_{\mathrm{succ}}$ is the success probability and $\overline{F}_{\mathrm{prep}}$ is the prepared-state fidelity of Eq.~(\ref{eq:Fprep}).}
\label{tab:midorder}
\begin{ruledtabular}
\begin{tabular}{lccc}
Model & $n$ & $p_{\mathrm{succ}}$ & $\overline{F}_{\mathrm{prep}}$ \\
\colrule
Spin-1 {\it XY} & 2 & $0.69$ & $0.73$ \\
Spin-1 {\it XY} & 3 & $0.63$ & $0.63$ \\
DWC & 2 & $0.71$ & $0.66$ \\
DWC & 3 & $0.68$ & $0.37$ \\
Onsager & 2 & $0.72$ & $0.66$ \\
Onsager & 3 & $0.70$ & $0.43$ \\
\end{tabular}
\end{ruledtabular}
\end{table}

\begin{figure}[htbp]
\centering
\includegraphics[width=0.72\linewidth]{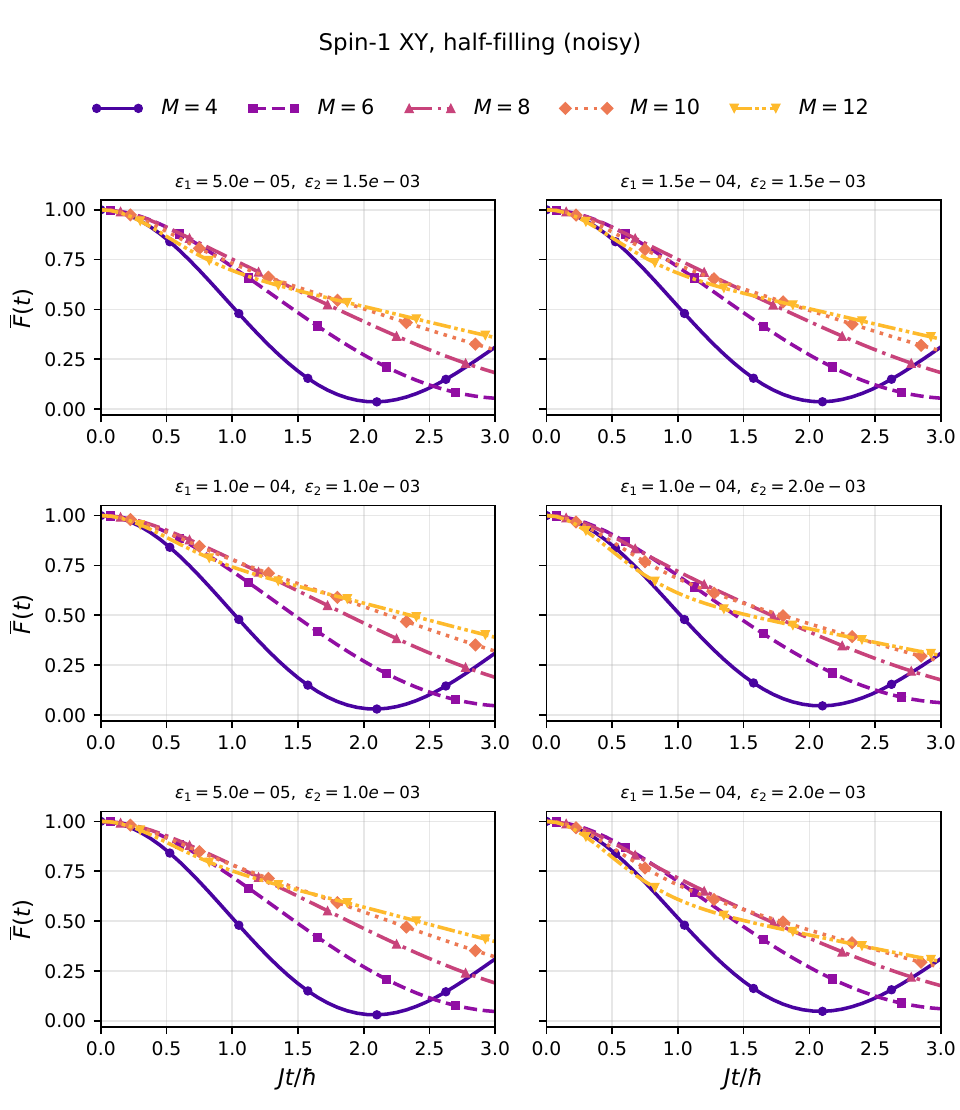}
\caption{Spin-1 {\it XY} model at half filling ($n=M/2$) under depolarizing gate noise, sizes $M=4,6,8,10,12$ for $l=1$. Each panel corresponds to a distinct one-/two-qubit error pair $(\varepsilon_1,\varepsilon_2)$, printed above it (the six selected non-central grid points; the central point is shown in Fig.~\mainref{fig:relax}{2}(d) in the main text).}
\label{fig:sm_spin1xy_half}
\end{figure}

Figure~\ref{fig:sm_spin1xy_half} shows the self-overlap fidelity of the spin-1 {\it XY} model at half filling across the error grid. For every error pair shown, the $M=12$ curve lies below those for $M=6$--$10$ at short times, as also observed at the central error rates [Fig.~\mainref{fig:relax}{2}(d) in the main text]. Across the scanned ranges, the $\overline{F}(t)$ curves are more sensitive to $\varepsilon_2$ than to $\varepsilon_1$. For $M=12$ at $Jt/\hbar=0.9$, doubling $\varepsilon_2$ from $10^{-3}$ reduces $\overline{F}(t)$ from about $0.77$ to $0.64$ at fixed $\varepsilon_1=10^{-4}$. By contrast, tripling $\varepsilon_1$ from $5\times10^{-5}$ reduces $\overline{F}(t)$ from about $0.72$ to $0.71$ at fixed $\varepsilon_2=1.5\times10^{-3}$. The DWC and Onsager scar models show a similar dependence on the error rates (Figs.~\ref{fig:sm_dwc_half} and~\ref{fig:sm_onsager_half}).

\begin{figure}[htbp]
\centering
\includegraphics[width=0.9\linewidth]{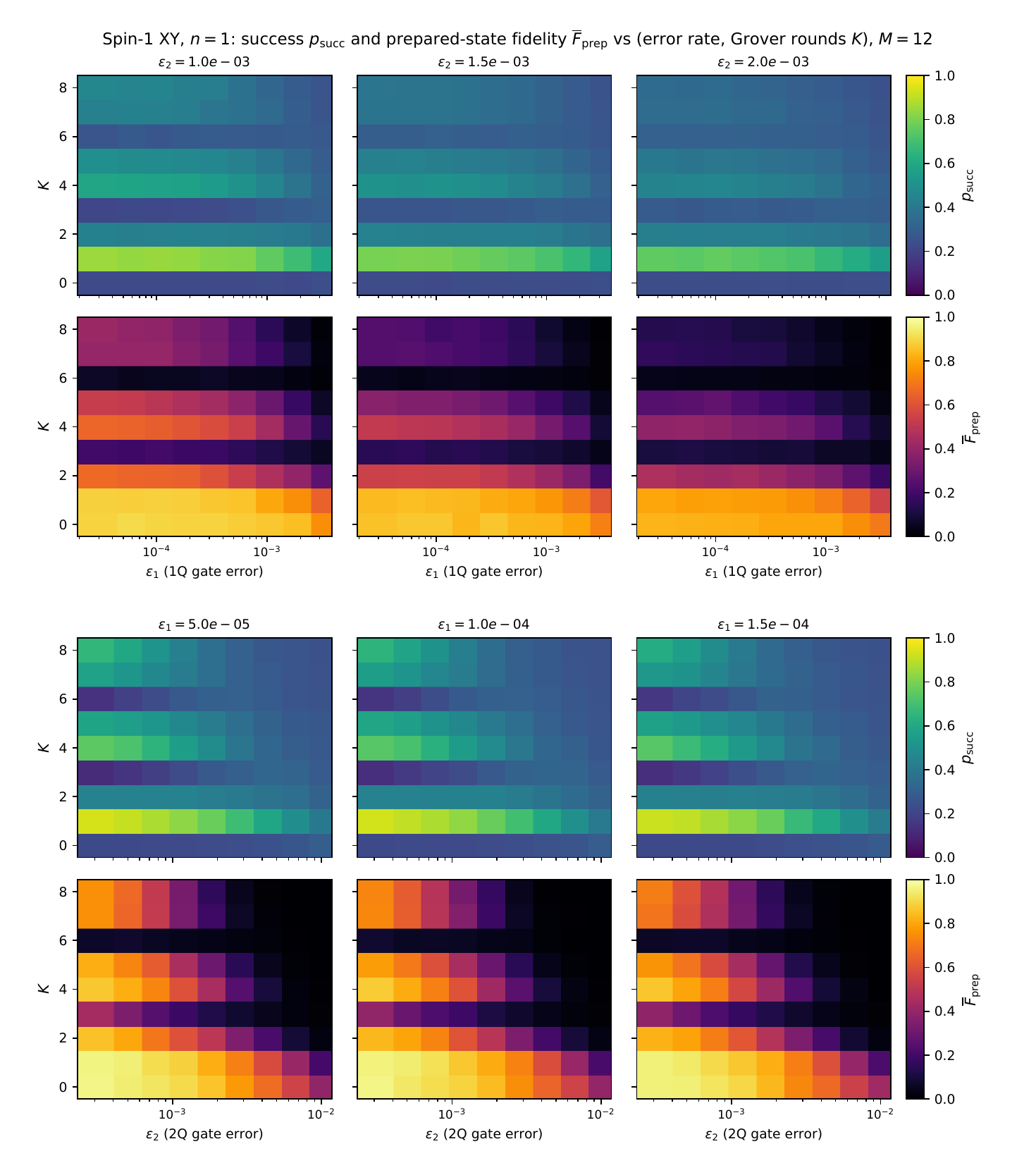}
\caption{Spin-1 {\it XY} model, scar index $n=1$, $M=12$. Success (postselection) probability $p_{\mathrm{succ}}$ and prepared-state fidelity $\overline{F}_{\mathrm{prep}}$ as functions of the number of Grover rounds $K$ and the gate error rate. Upper block: the one-qubit error rate $\varepsilon_1$ is swept at three fixed two-qubit rates $\varepsilon_2$ (column titles); lower block: $\varepsilon_2$ is swept at three fixed values of $\varepsilon_1$. Within each block the top row shows $p_{\mathrm{succ}}$ and the bottom row $\overline{F}_{\mathrm{prep}}$.}
\label{fig:sm_heat_spin1xy_n1}
\end{figure}

\begin{figure}[htbp]
\centering
\includegraphics[width=0.9\linewidth]{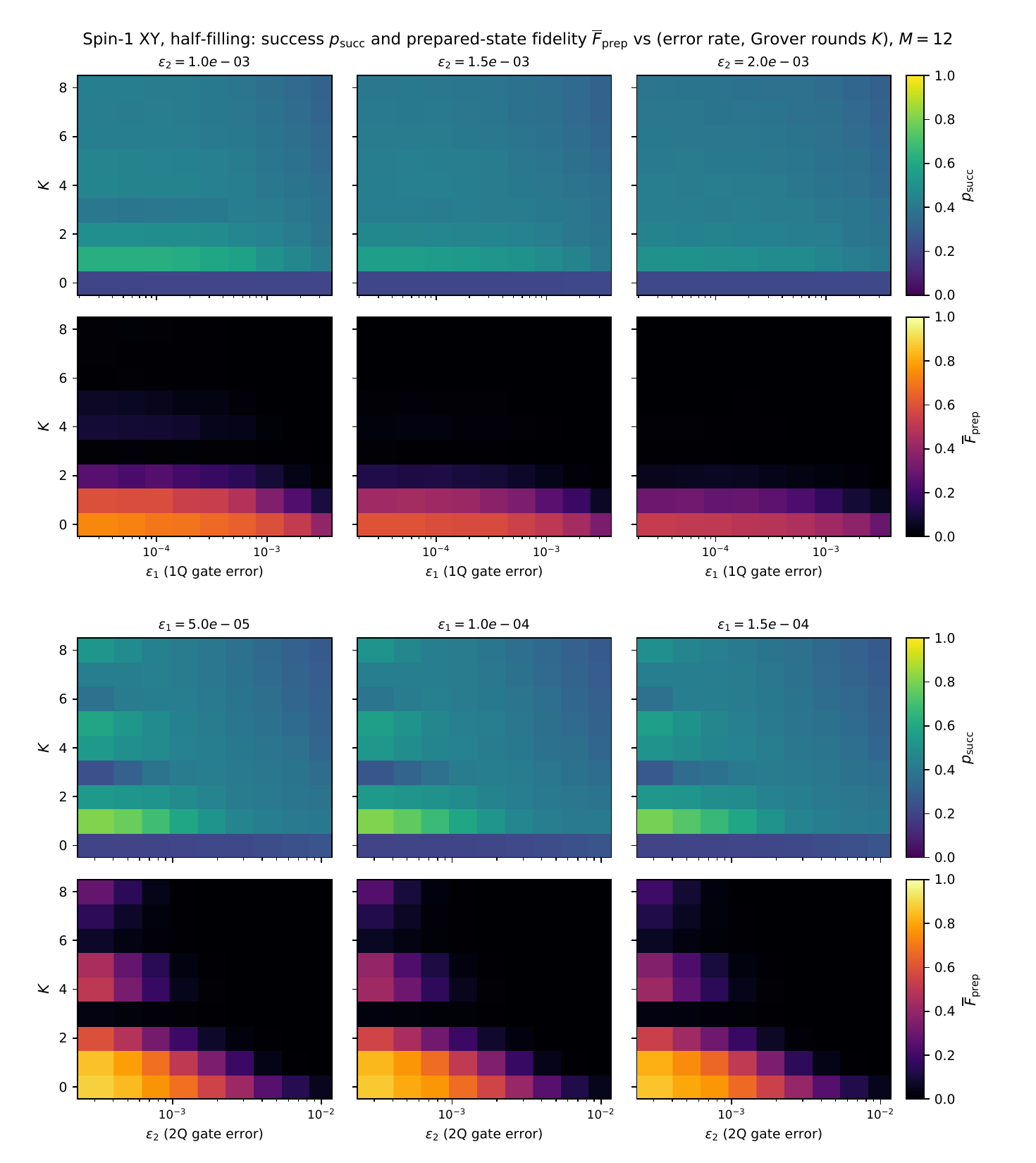}
\caption{Spin-1 {\it XY} model at half filling ($n=M/2$), $M=12$. Same layout as Fig.~\ref{fig:sm_heat_spin1xy_n1}.}
\label{fig:sm_heat_spin1xy_half}
\end{figure}

\begin{figure}[htbp]
\centering
\includegraphics[width=0.9\linewidth]{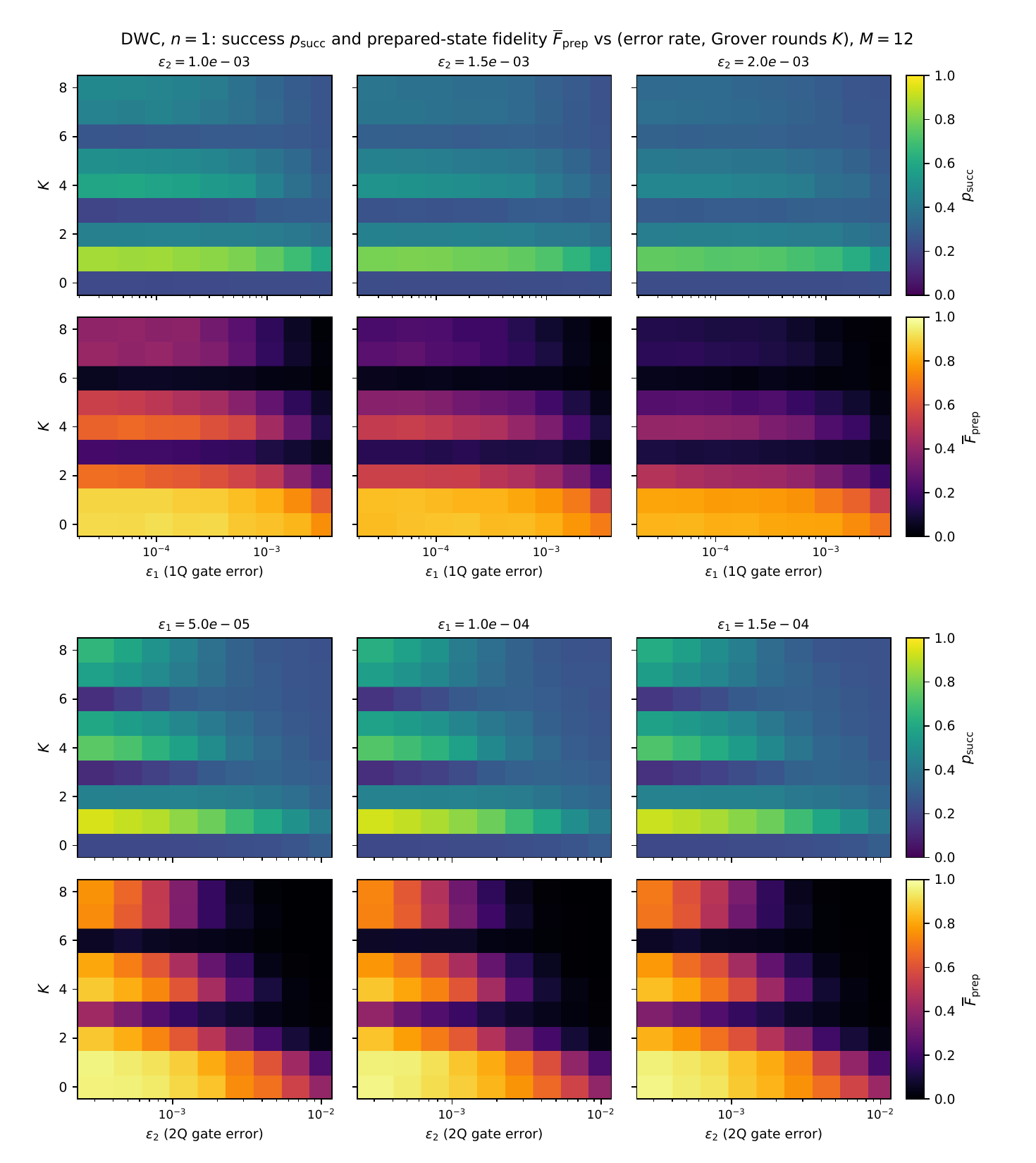}
\caption{DWC model, scar index $n=1$, $M=12$. Same layout as Fig.~\ref{fig:sm_heat_spin1xy_n1}.}
\label{fig:sm_heat_dwc_n1}
\end{figure}

\begin{figure}[htbp]
\centering
\includegraphics[width=0.9\linewidth]{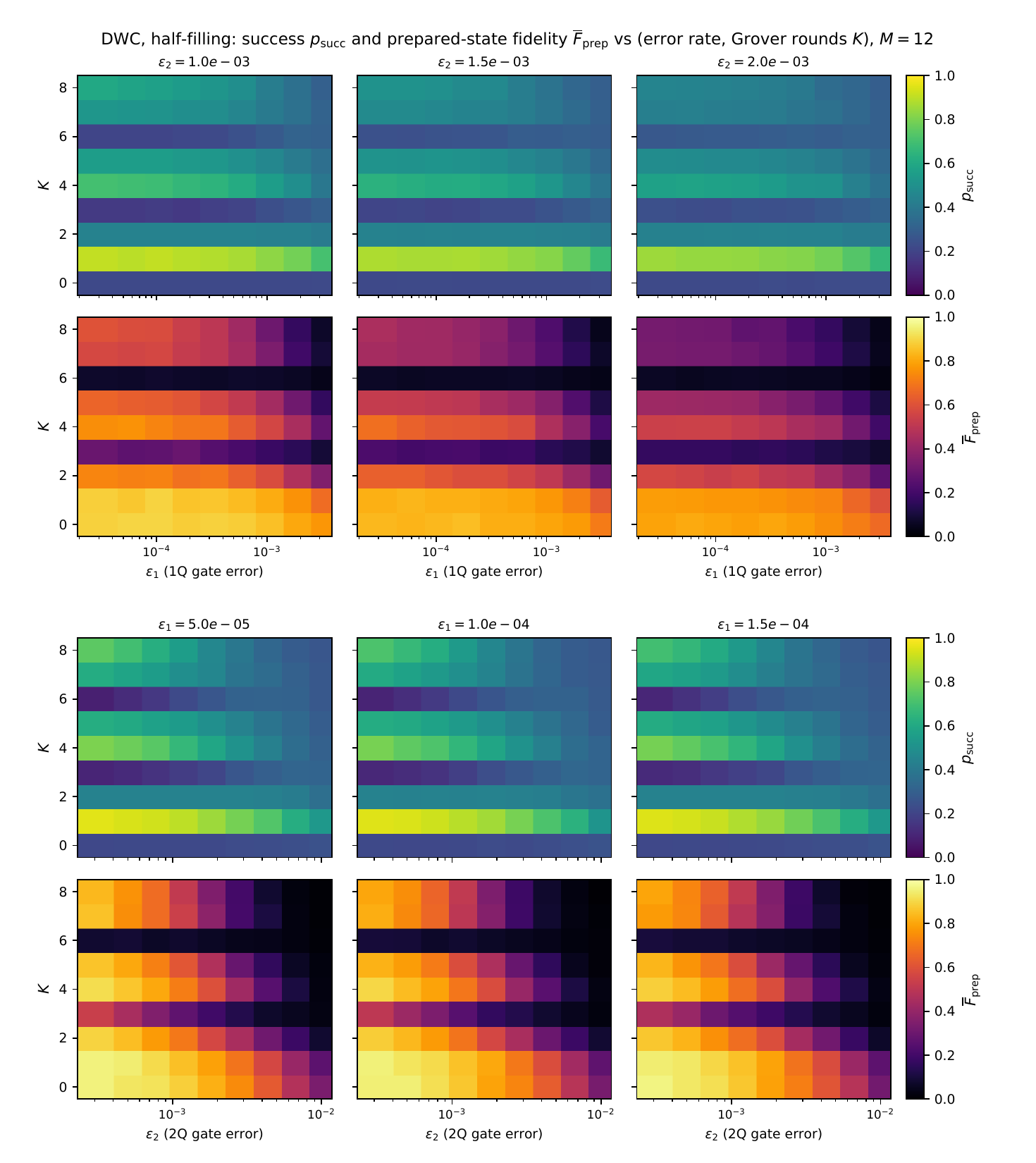}
\caption{DWC model at half filling ($n=M/2$), $M=12$. Same layout as Fig.~\ref{fig:sm_heat_spin1xy_n1}.}
\label{fig:sm_heat_dwc_half}
\end{figure}

\begin{figure}[htbp]
\centering
\includegraphics[width=0.9\linewidth]{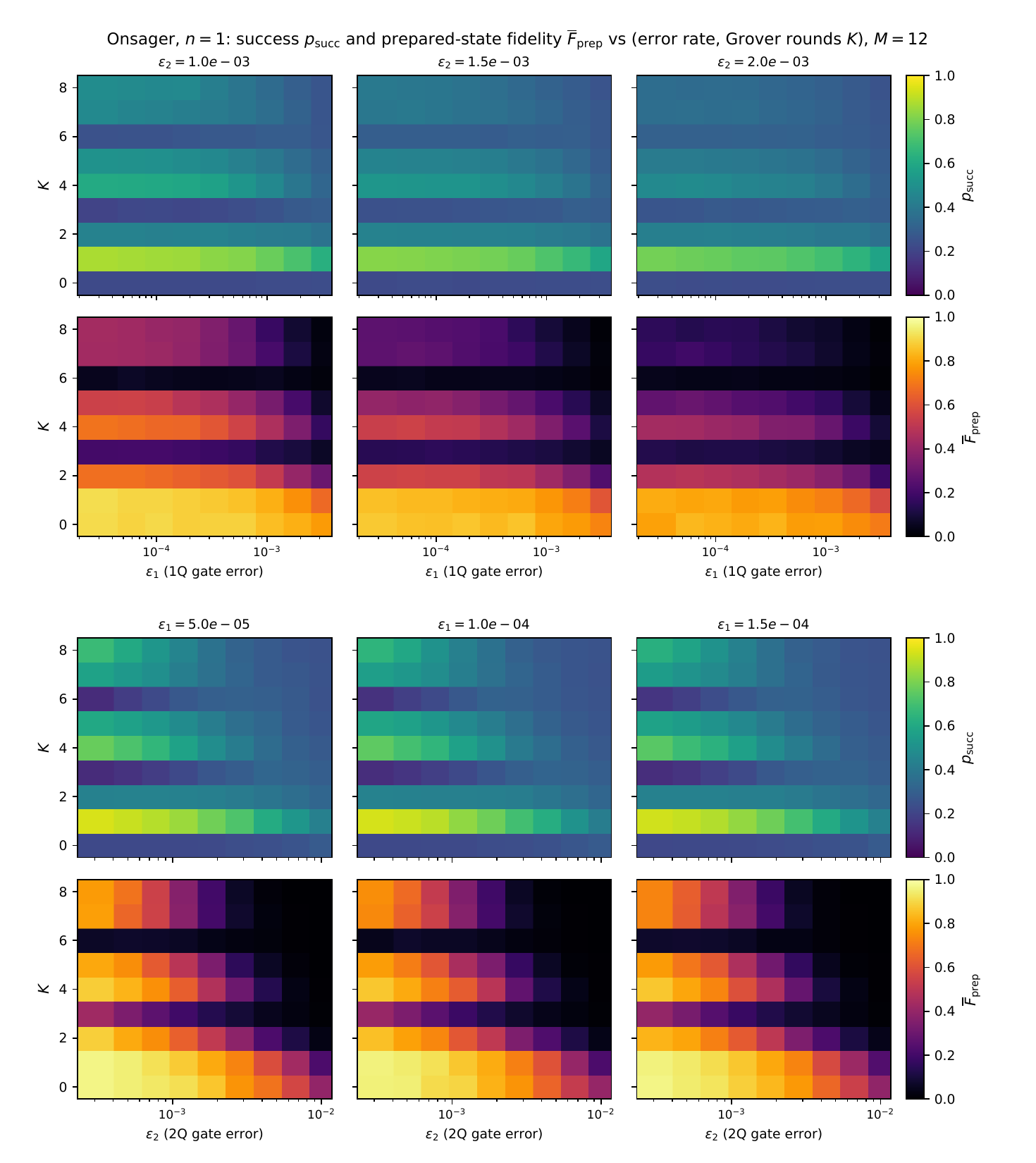}
\caption{Onsager scar model, $n=1$, $M=12$. Same layout as Fig.~\ref{fig:sm_heat_spin1xy_n1}.}
\label{fig:sm_heat_onsager_n1}
\end{figure}

\begin{figure}[htbp]
\centering
\includegraphics[width=0.9\linewidth]{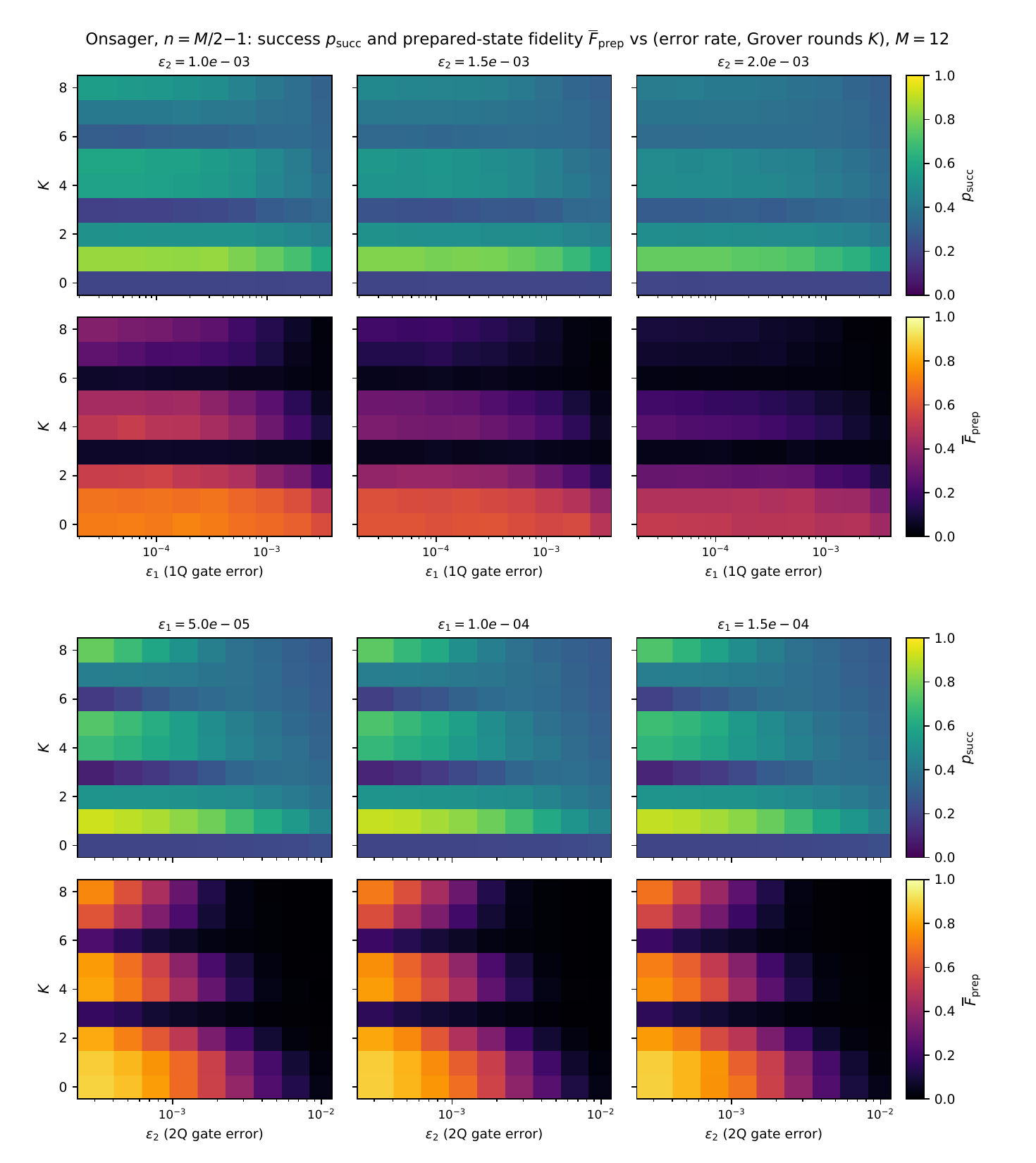}
\caption{Onsager scar model near the maximal scar index ($n=M/2-1$), $M=12$. Same layout as Fig.~\ref{fig:sm_heat_spin1xy_n1}.}
\label{fig:sm_heat_onsager_half}
\end{figure}

Figures \ref{fig:sm_heat_spin1xy_n1}--\ref{fig:sm_heat_onsager_half} show the success probability $p_{\mathrm{succ}} = \frac{1}{2400}\sum_{i=1}^{2400} p_{\mathrm{succ}, i}$ and the prepared-state fidelity
\begin{equation}
\label{eq:Fprep}
    \overline{F}_{\text{prep}} = \frac{\sum_i p_{\mathrm{succ},i} |\braket{AS_n (l)| \psi_i (0)}|^2}{\sum_i p_{\mathrm{succ},i}},
\end{equation}
as functions of the number of Grover rounds $K$ and the gate error rates. Here, $\ket{AS_n(l)}=\Phi^{-1}|A\rangle$ is the normalized ideal target state in the physical Hilbert space, where $\ket{A}$ is the normalized state defined in Eq.~\eqref{eq:ASdef}, $\ket{\psi_i (0)}$ is the postselected initial state of the $i$th trajectory and $p_{\mathrm{succ},i}$ is its success probability. The success probability $p_{\text{succ}, i}$ of the $i$th trajectory is the probability of finding the flag in $\ket{1}$ and the work ancilla of $R_0$ in $\ket{0}$.  The success probability oscillates with $K$ and peaks at $K=1$ for the cases studied at the central error rates. Meanwhile, $\overline{F}_{\mathrm{prep}}$ is more sensitive to two-qubit errors than to single-qubit errors. Without amplification ($K=0$), direct postselection succeeds with probability 0.21--0.23. Thus, a single amplification round provides a favorable trade-off between success probability and fidelity under the noise conditions considered.
\clearpage

\clearpage
\section*{D.\quad Entanglement of the prepared state}

We examine the half-chain entanglement entropy (EE) of the AQMBS states in the spin-1 {\it XY} model. Since the model conserves total magnetization $\tau^z_{\mathrm{tot}}\equiv \sum_j\tau_j^z$, the noise-free prepared state with scar index $n$ belongs to the sector $\tau^z_{\mathrm{tot}}=2n-M$. We use the Page value for Haar-random pure states in this target magnetization sector as a reference~\cite{BianchiDona2019,Yauk2024}.

For the half-chain bipartition with $L=M/2$ sites per half, let $m_q$ and $n_q$ denote the numbers of configurations with magnetizations $q$ and $2n-M-q$ in the two halves, respectively. The corresponding block has dimension $m_qn_q$, and the Page value is given exactly by~\cite{BianchiDona2019,Yauk2024}:
\begin{equation}
\label{eq:sector_page}
  S_{\mathrm{Page}}=\sum_{q}\frac{m_qn_q}{B}\left\{\Psi(B+1)-\Psi\left[\max(m_q,n_q)+1\right]-\frac{\min(m_q,n_q)-1}{2\max(m_q,n_q)}\right\},
  \qquad B\equiv \sum_{q}m_qn_q,
\end{equation}
where $\Psi(z)\equiv d\ln\Gamma(z)/dz$ is the digamma function, and $\Gamma(z)\equiv\int_0^\infty t^{z-1}e^{-t}dt$ is the gamma function ($z>0$). For a single block, Eq.~\eqref{eq:sector_page} reduces to Page's formula~\cite{Page1993}. Below we evaluate this expression for $n=1$ and $n=M/2$.

\begin{figure}[htbp]
\centering
\includegraphics[width=0.95\linewidth]{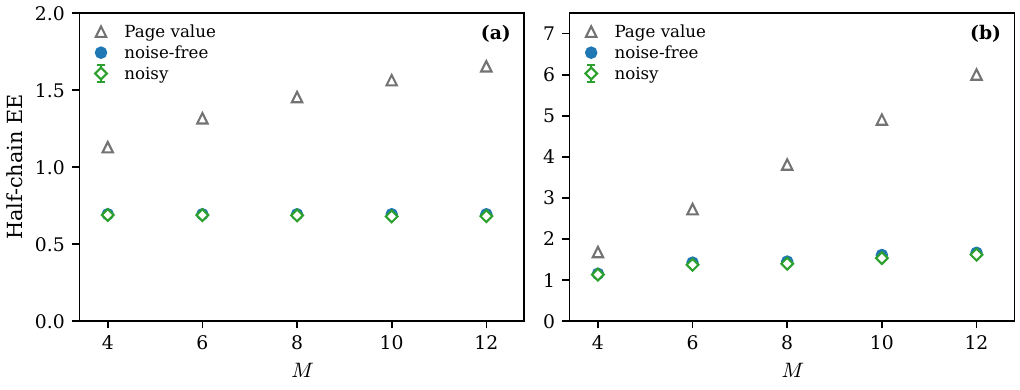}
\caption{
Half-chain entanglement entropy of the AQMBS states in the spin-1 {\it XY} model: (a) $n=1$ and (b) half filling ($n=M/2$). Gray triangles show the exact Page values in the corresponding magnetization sectors [Eqs.~\eqref{eq:page_n1} and \eqref{eq:page_half_exact}]. Blue circles show the noise-free EE, and green diamonds show the trajectory-averaged EE under gate noise at the central error rates. Error bars indicate one standard error and are smaller than the symbols.
}
\label{fig:sm_entanglement}
\end{figure}

For $n=1$, the two units of magnetization above the fully polarized state can lie entirely in either half or be split between them. The three blocks therefore have $(m_q,n_q)=(1,C_L)$, $(L,L)$, and $(C_L,1)$, where $C_L=L(L+1)/2$ and $B=L(2L+1)=M(M+1)/2$. Substituting $L=M/2$ into Eq.~\eqref{eq:sector_page} gives the exact expression
\begin{align}\label{eq:page_n1}
  S_{\mathrm{Page}}^{(n=1)}
  =&\Psi\left[\frac{M(M+1)}{2}+1\right]
  -\frac{M+2}{2(M+1)}
  \Psi\left[\frac{M(M+2)}{8}+1\right]\notag\\
  &-\frac{M}{2(M+1)}
  \Psi\left[\frac{M}{2}+1\right]
  -\frac{M-2}{4(M+1)}\notag\\
  =&\frac{M+2}{2(M+1)}\sum_{k=C_L+1}^{B}\frac{1}{k}+\frac{M}{2(M+1)}\sum_{k=M/2+1}^{B}\frac{1}{k}-\frac{M-2}{4(M+1)}.
\end{align}
In the zero magnetization sector ($\tau^z_{\mathrm{tot}}=0$), spin-inversion symmetry gives $n_q=m_q$. Equation~\eqref{eq:sector_page} then reduces to the exact expression
\begin{align}
  \label{eq:page_half_exact}
  S_{\mathrm{Page}}^{(n=M/2)}
  =&\Psi(B+1)
  -\sum_{q=-L}^{L}\frac{m_q^2}{B}
  \left[
    \Psi(m_q+1)+\frac{m_q-1}{2m_q}
  \right]\notag\\
  =&\sum_{q=-L}^{L}\frac{m_q^2}{B}
  \left[
    \sum_{k=m_q+1}^{B}\frac{1}{k}-\frac{m_q-1}{2m_q}
  \right],\\
  m_q=&
  \sum_{r=0}^{\lfloor(L-|q|)/2\rfloor}
  \frac{L!}
  {r!(r+|q|)!(L-2r-|q|)!},\\B=&\sum_{q=-L}^{L}m_q^2.
\end{align}
Figure~\ref{fig:sm_entanglement} compares the exact Page values from Eqs.~\eqref{eq:page_n1} and \eqref{eq:page_half_exact} with the EE of the prepared states. For $n=1$, reflection symmetry [$f_{M+1-j}(1)=-f_j(1)$] gives equal probabilities of finding the excitation in either half, yielding an EE of exactly $\ln2$. At half filling, the EE grows logarithmically with $M$. At $M=12$, the Page values exceed the noise-free prepared-state EE by factors of $2.4$ and $3.6$ for $n=1$ and half filling, respectively.

We define the trajectory-averaged EE as $\overline{S}_{\rm vN}\equiv\sum_j p_{{\rm succ},j}\,S_{{\rm vN},j}/\sum_j p_{{\rm succ},j}$, where $S_{{\rm vN},j}$ is the half-chain von Neumann EE of the state $|\psi_j(0)\rangle$ prepared in the $j$th trajectory. At the central gate error rates, the trajectory-averaged EE is $0.7~\%$--$4.6~\%$ below the noise-free values, with standard errors below $0.006$. Although each trajectory has a well-defined EE, its average depends on the unraveling of the noise channel and is therefore used only as a numerical diagnostic.

\clearpage
\bibliography{references_aqmbs}